\documentclass[11pt]{article}

\usepackage{cite,epsf,picinpar, graphicx}
\usepackage{latexsym}
\usepackage{amssymb}
\usepackage{amsmath}
\usepackage{makeidx}     
\usepackage{graphicx}    
\usepackage{multicol}    
\usepackage{epsfig}
\usepackage{longtable}
\usepackage{epstopdf}
\begin{document}

\renewcommand\arraystretch{1.8}
\def\p#1#2{\frac{\partial #1}{\partial #2}}
\def\f#1#2{\frac{#1}{#2}}
\def\ep{\epsilon}       \def\pa{\partial}
\def\ga{\gamma}         \def\Ga{\Gamma}
\def\del{\delta}        \def\Del{\Delta}
\def\si{\sigma}         \def\Si{\Sigma}
\def\om{\omega}         \def\Om{\Omega}
\def\bt{\beta}          \def\tr{\triangle} \def\na{\nabla}
\def\sb{\subset}        \def\sp{\supset}
\def\uv{(u,v)}          \def\tx{(t,x)}
\def\ts{(t,s)}          \def\vp{\varphi}
\def\ah{\alpha}         \def\u{\underline}
\def\b{\bar}            \def\l{\lambda}
\def\ue{\mathbf{u}_{\ep}}      \def\re{\rho_{\ep}}
\def\uu{\mathbf{u}}            \def\me{\mathbf{m}}
\def\vv{\mathbf{v}}      \def\xx{\mathbf{x}}
\def\R{{\mathbb R}}
\def\ml{\mathcal{L}}           \def\Pro#1{{\bf Proposition #1.}}
\def\Th#1{{\bf Theorem #1.}}
\def\Co#1{{\bf Conclusion #1.}}
\def\Lm#1{{\bf Lemma #1.}}
\def\Re#1{{\bf Remark #1.}}
\def\De#1{{\bf Definition #1.}}
\def\Pr{{\bf Proof \ \ }}
\def\be{\begin{array}{l}}
\def\ee{\end{array}}
\def\eem{\end{array}}
\def\ben{\begin{eqnarray}}
\def\een{\end{eqnarray}}
\def\bee{\begin{eqnarray*}}
\def\eee{\end{eqnarray*}}
\def\dv{\mathrm{div}\,}
\setlength{\arraycolsep}{0.5mm}

\title{\bf Unified gas-kinetic simulation for the system of multiscale radiation hydrodynamics}
\author{Wenjun Sun$^{\dag}$ \qquad Song Jiang$^\dag$ \qquad  Kun Xu$^{\ddag,}$\footnote{Corresponding author.}\qquad  Guiyu Cao$^{\ddag}$
 \\
{\small $^\dag$ Institute of Applied Physics and Computational Mathematics,  No.2, FengHao East Road, }\\[-4mm]
{\small HaiDian District, Beijing 100094, China.} \\[-2mm]
{\small $^\ddag$ Department of Mathematics, Hong Kong University of
Science and Technology,}\\[-4mm] {\small Clear Water Bay, Kowloon, Hong Kong.}\\[-2mm]
{\small E-mail: sun\_wenjun@iapcm.ac.cn, \ \  jiang@iapcm.ac.cn, \ \ makxu@ust.hk, \ \ gcaoaa@connect.ust.hk   } }
\date{}
\maketitle
\begin{abstract}
This paper aims at the simulation of multiple scale physics in the system of radiation hydrodynamics.
The system couples the fluid dynamic evolution equations with the radiative heat transfer.
The coupled system is solved by the gas-kinetic scheme (GKS) for the compressible viscous and heat conducting flow and the
unified gas-kinetic scheme (UGKS) for the non-equilibrium radiative transfer, together with the momentum and energy exchange
between these two phases.
For the radiative transfer, due to the possible large variation of fluid opacity in different regions,
the transport of photons in the flow system is simulated by the multiscale UGKS, which is capable of naturally  capturing
the transport process from the free streaming  to the diffusive propagation.
Since both GKS and UGKS are finite volume methods,
all unknowns are defined inside each control volume and are discretized consistently for the hydrodynamic and radiative variables.
For the coupled system, the scheme has the asymptotical preserving (AP) property, such as
       recovering the equilibrium diffusion limit for the radiation hydrodynamic equations in the optically thick region, where
        the cell size is not limited by photon's mean free path.
A few test cases, such as radiative shock wave problems, are used to validate the current approach.
\end{abstract}


\renewcommand{\theequation}{\thesection.\arabic{equation}}
\setcounter{equation}{0}
\section{Introduction}

This paper aims at constructing an asymptotic preserving numerical scheme for radiation hydrodynamics.
Radiation hydrodynamics describes radiative transport through a fluid with coupled momentum and energy exchange.
It is popularly used in high energy density physics, astrophysics, the inertial confinement fusion (ICF), and other flows
with very high temperatures.
For radiation hydrodynamics, radiation is propagating through a moving hydrodynamic material with the coupled momentum
and energy exchange. Due to the material velocity, the thermal radiative transfer equation requires a certain amount of
material-motion correction whenever the radiation momentum deposition has a measurable impact on the
material dynamics. This is even true for flows with material speed being much smaller than the
speed of light. Following the works of \cite{MM1, LMH1, LW1}, we adopt Morel's radiation hydrodynamic model in this paper,
denoted as $MM(\theta)$ model in equation (\ref{Ap2.1}) and $\theta$ as a free parameter.
Morel's system can be viewed as a simplified laboratory-frame formulation. The parameter $\theta$ can be chosen based on
the numerical method, such as $MM(\theta = 1)$ for a Lagrangian approach and $MM(\theta = 0)$ for the Eulerian one.
Based on Morel's model and considering the multiple time scales in the radiation hydrodynamic equations,
the equations (\ref{Ap2.1}) are usually split into the small time scale part of radiation
and large time scale part of of the fluid.

Though detailed work has been done individually for time integration of radiative
transfer \cite{MELD, Lowrie1, KLM1, Olson1, ADJ1, SM1, BSW1, SJXu, SJX2, SJX4} and for fluid dynamics \cite{T1, Xu1},
the research on the coupled system has only been carried out recently \cite{KKLM, BHEM, Lowrie, LW1, DW1, BKRLM}.
The equations of radiation hydrodynamics include explicitly the motion of the background material.
For low opacity material, such as the case with small absorption/emission coefficients and small scattering
coefficient, the interaction between the radiation and material is
weak and the radiation propagates in a transparent way with the particle-type behavior, i.e., the so-called optically thin regime.
In this regime, the numerical method for radiation should be able to capture the streaming transport of photons,
such as the upwind approach with a ray tracking technique in SN method.
For a high opacity material with large absorption/emission coefficients or large scattering coefficient,
the intensive momentum and energy exchange between the radiation and material makes photon's mean free path diminish.
As a result, different asymptotic limits in the optically thick regime will appear.
In the case with large absorption/emission coefficients, an equilibrium diffusive process for radiation will emerge and
the material temperature and the radiation temperature will get the same value.
In this paper, the unified gas-kinetic scheme (UGKS) will be used for the radiative transfer part to capture both ballistic
and diffusive limits for the photon transport \cite{SJXu,SJX2, SJX3, SJX4}.

For hydrodynamics, the gas-kinetic scheme (GKS) has been developed systematically for compressible flow computations \cite{Xu1, Xu2, Xu3}.
The numerical flux in the finite volume GKS is constructed based on a gas evolution process from a kinetic scale particle free transport
to a hydrodynamic scale Navier-Stokes flux formulation, where both inviscid and viscous fluxes are recovered from moments of a single
time-dependent gas distribution function. In the discontinuous shock region, the GKS becomes a shock capturing scheme and the
kinetic scale based particle free transport, or so-called upwinding, takes effect to build a crisp
and stable numerical shock transition. The highly non-equilibrium of the gas distribution function in
the discontinuous region provides a physically consistent mechanism for the construction of a numerical shock structure.

In this paper we construct a scheme for the radiation hydrodynamic system by coupling UGKS with GKS uniformly in all regimes.
Since both GKS and UGKS are finite volume method, all flow and radiation variables are defined as cell averages.
The discretization for both hydrodynamics and radiative evolution can be done consistently.
The constructed coupling scheme has the asymptotic preserving (AP) property for the radiative part,
where the equilibrium diffusion limit of radiation will be obtained automatically by UGKS in the optically thick region.

The plan of the paper is as follows. In Section \ref{sec:RHequ}, the radiation hydrodynamic system is presented.
In Section \ref{sec:solver}, the details of the numerical scheme for the coupled system are given.
In Section \ref{sec:APanal}, the asymptotic preserving property of the scheme is proved mathematically.
The numerical examples are presented in Section \ref{sec:results} to test the performance of the current scheme. Finally,
a conclusion is given in Section \ref{sec:con}.

\setcounter{equation}{0}
\section{Radiation hydrodynamics}\label{sec:RHequ}

For radiation hydrodynamics, when the radiation momentum deposit has a measurable impact on the material dynamics, the thermal radiative transfer equation requires the correction due to the material velocity. The modification is needed even for the case where the speed of flow is much smaller than the speed of light. Under such a condition,  the $MM(\theta)$ model \cite{LW1} for the coupled radiation and hydrodynamics will be adopted in the current study. The equations include  non-relativistic, inviscid, single-material compressible hydrodynamics and the
thermal radiation transport,
\begin{equation}
\left\{\be
\partial_t\rho + \nabla \cdot (\rho \vec{v}) = 0, \\[2mm]
\partial_t(\rho\vec{v}) + \nabla \cdot (\rho \vec{v}\otimes\vec{v}) + \nabla p = -\frac{1}{c}\displaystyle{\int\vec{\Omega}Sd\vec{\Omega}}, \\[2mm]
\partial_t(\rho E) + \nabla \cdot [\vec{v}(\rho E+p)] = -\frac{1}{\epsilon}\displaystyle{\int Sd\vec{\Omega}}, \\[2mm]
\displaystyle{\frac{\epsilon}{c}\frac{\partial I}{\partial t} }
+\vec{\Omega}\cdot \nabla I+\epsilon \nabla \cdot (\theta \vec{\beta} I) = - \frac{\sigma_t}{\epsilon}I + ( \frac{\sigma_t}{\epsilon}-\epsilon \sigma_s) \frac{1}{4\pi}acT^4 + \frac{\epsilon \sigma_s }{4\pi}cE_r\\[2mm]
\qquad\qquad\qquad -\frac{1}{4\pi}\sigma_t\vec{\beta}\cdot [\vec{F}_r-(\frac{4}{3}-\theta)\epsilon E_r\vec{v} ]+\frac{3}{4\pi}(\frac{4}{3}-\theta)\sigma_tE_r\vec{\Omega}\cdot\vec{v}\triangleq S. \ee \right.\label{Ap2.1}
\end{equation}
Here $\rho$ is the mass density, $T$ the material temperature, $\vec{v}$ the fluid velocity, and $\rho E=\frac{1}{2}\rho|\vec{v}|^2+\rho e$
is the total material energy. In order to close the equations, the equation-of-state (EOS) $p=p(\rho, T)$
and the material internal energy $e(\rho, T)$ have to be provided. And $I$ is the radiation intensity, which is a function of space, time,
angle direction $\vec{\Omega}$, and radiation frequency.
For simplicity, in this paper we only consider the gray case, where the intensity is averaged over the radiation frequency.
In the above equations, $c$ is the speed of light and $\vec{\beta}\equiv\frac{\vec{v}}{c}$.
The $S$ term represents the interaction between the radiation and material in the radiation hydrodynamic system,
$a$ is the radiation constant, $\sigma_s$ is the coefficient of scattering, $\sigma_t$ is the total coefficient of absorption,
and the $\epsilon$ is the factor of scaling. The free parameter $\theta$ is related to the correction due to the material motion.
The value of $\theta$ varies according to the numerical scheme. For the Lagrangian formulation with moving mesh following the fluid velocity,
  $\theta =1$ is used. In the Eulerian formulation, $\theta=0$ is adopted for the lab-frame,
 while the case $\theta=4/3$ can be viewed as an approximate comoving-frame treatment.
The functions $E_r$ and $\vec{F}_r$ are the radiation energy and radiation flux respectively, which are given by
$$ E_r = \frac{1}{c}\int Id\vec{\Omega},\qquad \vec{F}_r = \int \vec{\Omega}Id\vec{\Omega}. $$

The momentum and energy deposition from radiation on hydrodynamics are computed by angle integrals on the right hand sides
of the second and third equations in (\ref{Ap2.1}). It is straightforward to derive the corresponding total momentum and energy equations,
which are given by
 \begin{equation}
\left\{\be
\partial_t(\rho \vec{v} + \frac{\epsilon}{c^2}\vec{F}_r) + \nabla \cdot (\rho \vec{v}\otimes\vec{v}+\frac{\epsilon\theta}{c^2}\vec{v}\otimes\vec{F}_r+\bar{\bar{P}}) +\nabla p= 0, \\[2mm]
\partial_t(\rho E + E_r ) + \nabla \cdot [ \vec{v}(\rho E + \theta E_r + p) + \frac{1}{\epsilon}\vec{F}_r ] = 0,
 \ee \right.\label{Ap2.2}
\end{equation}
and $\bar{\bar{P}}$ is the radiation pressure tensor calculated by
$$ \bar{\bar{P}} = \frac{1}{c}\int \vec{\Omega}\otimes\vec{\Omega}I d\vec{\Omega}. $$

The system (\ref{Ap2.1}) has the property that it will approach to the equilibrium diffusion limit equations
for any choice of $\theta$ as the parameter $\epsilon$ approaching to $0$ in the optically thick region.
This can be seen by expanding the dependent variables as a power series of $\epsilon$,
\begin{equation}
\left\{\be
\rho = \displaystyle{\sum_{i=1}^{\infty}}\rho^{(i)}\epsilon^i, \ \ \ \ \ \
\vec{v} = \displaystyle{\sum_{i=1}^{\infty}}\vec{v}^{(i)}\epsilon^i, \\[2mm]
T = \displaystyle{\sum_{i=1}^{\infty}}T^{(i)}\epsilon^i, \ \ \ \ \ \
I = \displaystyle{\sum_{i=1}^{\infty}}I^{(i)}\epsilon^i,
 \ee \right.\label{Ap2.3}
\end{equation}
and comparing the terms of equal powers. Substituting the expansions in (\ref{Ap2.3}) into the governing equations (\ref{Ap2.1}),
the $O(\epsilon^{-1})-$terms of the fourth equation in (\ref{Ap2.1}) give
\begin{equation} I^{(0)} = \frac{1}{4\pi}ac(T^{(0)})^4, \label{Ap2.4}\end{equation}
followed by
\begin{equation}E_r^{(0)} = a(T^{(0)})^4, \ \ \vec{F}_r^{(0)} = 0, \ \  \bar{\bar{P}}^{(0)} = \frac{1}{3}a(T^{(0)})^4\bar{\bar{\textbf{D}}}, \label{Ap2.5}\end{equation}
where $\bar{\bar{\textbf{D}}}$ is the identity matrix.
There are no $O(\epsilon^{-1})-$terms in the first two equations of (\ref{Ap2.1}). And the $O(\epsilon^{-2})$ and $O(\epsilon^{-1})-$terms
in the third equation of (\ref{Ap2.1}) are consistent with the above Eqs. (\ref{Ap2.4}) and (\ref{Ap2.5}).

Using Eqs. (\ref{Ap2.4}) and (\ref{Ap2.5}) again, the $O(\epsilon^{0})-$terms in the fourth equation of (\ref{Ap2.1}) reduce to
\begin{equation}I^{(1)} =  \frac{1}{4\pi}ac(T^{(1)})^4 - \frac{c}{\sigma_t^{(0)}}\vec{\Omega} \cdot \nabla I^{(0)} + \frac{3}{4\pi}(\frac{4}{3}-\theta)E_r^{(0)}\vec{\Omega} \cdot\vec{v}^{(0)}, \label{Ap2.6}\end{equation}
therefore,
\begin{equation}E_r^{(1)} = a(T^{(1)})^4, \ \ \vec{F}_r^{(1)} = -\frac{c}{3\sigma_t^{(0)}}\nabla E_r^{(0)} + (\frac{4}{3}-\theta)E_r^{(0)}\vec{v}^{(0)},  \ \  \bar{\bar{P}}^{(1)} = \frac{1}{3}a(T^{(1)})^4\bar{\bar{\textbf{D}}}.\label{Ap2.7}\end{equation}
Finally, the $O(\epsilon^{0})-$terms in the first equation of (\ref{Ap2.1}) and the equations (\ref{Ap2.2}) result in
\begin{equation}
\left\{\be
\partial_t\rho^{(0)} + \nabla \cdot (\rho^{(0)} \vec{v}^{(0)}) = 0, \\[2mm]
\partial_t(\rho^{(0)} \vec{v}^{(0)}) + \nabla \cdot (\rho^{(0)} \vec{v}^{(0)}\otimes\vec{v}^{(0)}+\bar{\bar{P}}^{(0)}) +\nabla p^{(0)}= 0, \\[2mm]
\partial_t(\rho^{(0)} E^{(0)} + E_r^{(0)} ) + \nabla \cdot [ \vec{v}^{(0)}(\rho^{(0)} E^{(0)} + \theta E_r^{(0)} + p^{(0)}) + \vec{F}_r^{(1)} ] = 0.
 \ee \right.\label{Ap2.8}
\end{equation}

By substituting $\vec{F}_r^{(1)}$ in (\ref{Ap2.7}) into (\ref{Ap2.8}),
the equilibrium diffusion system for radiation hydrodynamics can be obtained as follows.
\begin{equation}
\left\{\be
\partial_t\rho^{(0)} + \nabla \cdot (\rho^{(0)} \vec{v}^{(0)}) = 0, \\[2mm]
\partial_t(\rho^{(0)} \vec{v}^{(0)}) + \nabla \cdot (\rho^{(0)} \vec{v}^{(0)}\otimes\vec{v}^{(0)}+\bar{\bar{P}}^{(0)}) +\nabla p^{(0)}= 0, \\[2mm]
\partial_t(\rho^{(0)} E^{(0)} + E_r^{(0)} ) + \nabla \cdot [ \vec{v}^{(0)}(\rho^{(0)} E^{(0)} + \frac{4}{3} E_r^{(0)} + p^{(0)})] = \nabla \cdot( \frac{c}{3\sigma_R}\nabla E_r^{(0)} ),
 \ee \right.\label{Ap2.9}
\end{equation}
where $\sigma_R$ is the Rosseland mean that is equal to $\sigma_t^{(0)}$ here.

This paper will present a scheme with the asymptotic preserving property for the radiation hydrodynamic equations (\ref{Ap2.1}),
such that the numerical scheme for (\ref{Ap2.1}) will converge to a proper numerical method for (\ref{Ap2.9}) automatically
as the parameter $\epsilon$ tends to zero. The details of the method will be presented in the next section.

\setcounter{equation}{0}
\section{Unified scheme for the radiation hydrodynamic system}\label{sec:solver}

In this subsection we introduce the detailed construction of an asymptotic preserving scheme for (\ref{Ap2.1}).
The radiation and fluid parts in Eqs.(\ref{Ap2.1}) will be solved separately. For the fluid dynamics, the gas kinetic scheme (GKS)
as a Navier-Stokes (NS) flow solver is used, while the multiscale unified gas-kinetic scheme (UGKS) \cite{SJXu} is employed for the radiative transfer,
where two solvers are coupled in the momentum and energy exchanges. Since GKS and UGKS are all finite volume methods,
and all unknowns are defined inside each control volume, and the discretizations for the hydrodynamics and radiative transfer can be done consistently.
%

The hydrodynamic and radiative transfer solvers are based on the operator-splitting approach.
The purely hydrodynamic part of our scheme targets on the following Euler equations, even though the GKS is intrinsically a NS solver,
\begin{equation}
\left\{ \begin{array}{l}
\partial_t \rho + \nabla \cdot (\rho \vec{v} ) = 0, \\[2mm]
\partial_t ( \rho \vec{v} ) + \nabla \cdot (\rho \vec{v}\bigotimes\vec{v} ) + \nabla p = 0, \\[2mm]
\partial_t (\rho E ) + \nabla \cdot (\vec{v}(\rho E +p ) ) = 0. \end{array} \right.   \label{AP3.1}
\end{equation}
The above equations are closed by an ideal gas equation of state (EOS) and internal energy equation:
\begin{equation}
\left\{ \begin{array}{l}
p = (\gamma -1) \rho e, \\[2mm]
e = C_v T, \end{array} \right.   \label{AP3.2}
\end{equation}
where $\gamma$ is the specific heat ratio and $C_v$ is the heat capacity.

For the radiative transfer, the  momentum deposition and energy exchange between radiation and material are included in the coupled equations.
The algorithm for radiative transfer solves the following equations:
\begin{equation}
\left\{\be
\partial_t(\rho\vec{v})  = -\frac{1}{c}\displaystyle{\int\vec{\Omega} S d\vec{\Omega}}=\frac{\sigma_t}{\epsilon c}[\vec{F}_r-(\frac{4}{3}-\theta)\epsilon E_r\vec{v}  ], \\[2mm]
\partial_t(\rho E) = -\frac{1}{\epsilon}\displaystyle{\int S d\vec{\Omega}}=\frac{1}{\epsilon}( \frac{\sigma_t}{\epsilon}-\epsilon \sigma_s)(cE_r-acT^4)+\frac{\sigma_t}{\epsilon}\vec{\beta}\cdot [\vec{F}_r-(\frac{4}{3}-\theta)\epsilon E_r\vec{v}] , \\[2mm]
\displaystyle{\frac{\epsilon}{c}\frac{\partial I}{\partial t} }
+\vec{\Omega}\cdot \nabla I+\epsilon \nabla \cdot (\theta \vec{\beta} I) = - \frac{\sigma_t}{\epsilon}I + ( \frac{\sigma_t}{\epsilon}-\epsilon \sigma_s) \frac{1}{4\pi}acT^4 + \frac{\epsilon \sigma_s }{4\pi}cE_r\\[2mm]
\qquad\qquad-\frac{1}{4\pi}\sigma_t\vec{\beta}\cdot [\vec{F}_r-(\frac{4}{3}-\theta)\epsilon E_r\vec{v}  ]+\frac{3}{4\pi}(\frac{4}{3}-\theta)\sigma_tE_r\vec{\Omega}\cdot\vec{v}\triangleq S. \ee \right.\label{Ap3.3}
\end{equation}
The solver for the radiative hydrodynamic system is constructed by solving the equations (\ref{AP3.1}) and (\ref{Ap3.3}) by GKS and UGKS separately.

\subsection{Gas-kinetic scheme based fluid solver}

The compressible Euler equations (\ref{AP3.1}) is solved by the GKS \cite{Xu1}.
In the finite volume GKS, the interface flux between neighboring cells plays a dominant role for the quality of the scheme.
The gas evolution at a cell interface is constructed based on the following kinetic model equation \cite{bgk}:
\begin{equation}
f_t + \vec{u}\cdot \nabla f = \frac{g-f}{\tau},\label{AP3.4}
\end{equation}
where $f({\vec x},t,{\vec u})$ is the gas distribution function and $\vec u$ is the particle velocity.
The function $g$ is the equilibrium state approached by $f$ through a particle collision time $\tau$.
The collision term satisfies the compatibility condition
\begin{equation}
\int\frac{g-f}{\tau} \psi d\Xi = 0,\label{AP3.5}
\end{equation}
where $\psi =(1, \vec{u}, \frac{1}{2}(|\vec{u}|^2+|\vec{\xi}|^2) )^T$ is the collision invariants, $d\Xi=d\vec{u}d\vec{\xi}$,
and $\vec{\xi}=(\xi_1, \cdots, \xi_K)$ is the internal variable.

The connections between the macro quantities $(\rho,\rho\vec{v},\rho E)$ and their fluxes with the gas distribution function $f$ are given by
\begin{equation}
\be
\left( \be
\rho  \\
\rho \vec{v} \\
\rho E \ee \right ) =  \int \psi f d\Xi ,  \ \ \ \
\left( \be
\nabla \cdot (\rho \vec{v}) \\
\nabla \cdot (\rho \vec{v}\bigotimes\vec{v} ) +\nabla p  \\
\nabla \cdot [(\rho E + p) \vec{v}] \ee \right)  =  \int \psi \vec{u}\cdot \nabla f d\Xi.
\ee  \label{AP3.6}
\end{equation}
Once the gas distribution $f$ at a cell interface is fully determined, the numerical fluxes can be obtained.
In GKS, the boundary distribution function $f$ is evaluated from the integral solution of kinetic model equation (\ref{AP3.4}):
\begin{equation}
f(\vec{x}, t, \vec{u}, \vec{\xi} ) = \frac{1}{\tau}\int_0^tg(\vec{x}-\vec{u}(t-t'), t', \vec{u}, \vec{\xi})e^{-(t-t')/\tau}dt'
+  e^{-t/\tau}f_0(\vec{x}-\vec{u}t,\vec{u},\vec{\xi}).   \label{AP3.7}
\end{equation}
The initial condition $f_0$ in the above solution is modeled by
$$f_0 = f_0^l(\vec{x},\vec{u},\vec{\xi})  {\mathrm H} (( \vec{x} - \vec{x}_{s})\cdot\vec{n})
+ f_0^r(\vec{x},\vec{u},\vec{\xi})( 1 - {\mathrm H} ((\vec{x}-\vec{x}_s)\cdot\vec{n}) ), $$
where $\mathrm H$ is the Heaviside function, $f_0^l$ and $f_0^r$ are the initial gas distribution functions at the left and right sides of
a cell interface with a normal direction $\vec{n}$, and $\vec{x}_s$ is the center of the cell interface.
To keep a second-order accuracy, the initial distribution $f_0$ in space around $\vec{x}_s$ is approximated by piecewise polynomials
$$f_0^{l,r}(\vec{x},\vec{u},\vec{\xi}) = f_0^{l,r}(\vec{x}_s, \vec{u},\vec{\xi}) + (\vec{x}-\vec{x}_x)\cdot \nabla f_0^{l,r}(\vec{x}_s, \vec{u},\vec{\xi}).$$
Without loss of generality, with the assumption of $\vec{x}_s=0$, for the Euler solution (\ref{AP3.1})
the initial distribution functions $f_0^{l,r}(0)$ can be expressed as the Maxwellians,
$$ f_0^{l,r}(0)=g_0^{l,r}.$$
The equilibrium distribution functions $g_0^{l,r}$ are
$$g_0^{l,r} = \rho^{l,r} (\frac{{\lambda}^{l,r}}{\pi})^{\frac{K+2}{2}}e^{{\lambda}^{l,r}(|\vec{u}-{\vec{v}}^{l,r}|^2+|\vec{\xi}|^2)},$$
which are determined from the distributions of initial macroscopic flow variables $W^l=(\rho^l, (\rho\vec{v})^l, (\rho E)^l)$ and $W^r=(\rho^r, (\rho \vec{v})^r, (\rho E)^r)$.
The derivatives $\nabla f_0^{l,r}$, such as in the $x_k$-direction, are obtained from
\begin{equation}
\be
\left( \be
\frac{\partial \rho }{\partial x_k}|_{l,r} \\
\frac{\partial (\rho \vec{v})}{\partial x_k}|_{l,r} \\
\frac{\partial (\rho E) }{\partial x_k}|_{l,r} \ee \right ) = \int \psi \frac{\partial f_0^{l,r} }{\partial x_k}d\Xi, \ \ \ \ (k=1,\cdots, 3),
\ee  \label{AP3.7.1}
\end{equation}
where the derivatives of the macroscopic variables $(\frac{\partial \rho }{\partial x_k}|_{l,r}, \frac{\partial (\rho \vec{v})}{\partial x_k}|_{l,r},
  \frac{\partial (\rho E) }{\partial x_k}|_{l,r}$ are reconstructed with the MUSCL slope limiter \cite{vL}.

After determining the initial distribution function $f_0$, the equilibrium state $g$ in the integral solution (\ref{AP3.7}) can be expanded in space and time as
\begin{equation}
g = \bar{g} + \nabla \bar{g} \cdot \vec{x} + \frac{\partial \bar{g}}{\partial t}t, \label{AP3.8}
\end{equation}
where $\bar{g}$ is the equilibrium distribution function at a cell interface and is determined by the compatibility condition
$$\int\psi \bar{g}d\Xi = \bar{W} \triangleq (\bar{\rho},\bar{\rho}\vec{\bar{v}}, \bar{\rho} \bar{E})^T = \int_{\vec{u}\cdot\vec{n}>0}\psi g_0^l d\Xi+\int_{\vec{u}\cdot\vec{n}<0}\psi g_0^r d\Xi. $$

With the following notations
$$a_k^{l.r}={\bar{g}}_{x_k}^{l,r}/{\bar{g}}, \quad A^{l,r} = {\bar{g}}_t^{l,r}/{\bar{g}},$$
the spatial derivatives ${\bar{g}}_{x_k}^{l.r} = ({\partial \bar{g}}/{\partial x_k})|_{l,r}( k = 1, \cdots, 3)$
and time derivative ${\bar{g}}_t^{l,r} = ({\partial \bar{g}}/{\partial t})|_{l,r}$ are obtained from
the relations
$$\int\psi a_k^{l.r} d\Xi = \frac{\partial {\bar{W}}}{\partial x_k}|_{l,r}, \qquad \int\psi (\sum_{k=1}^{k=3}u_ka_k^{l,r}+A^{l,r})d\Xi = 0. $$
The derivatives for the macroscopic variables for the equilibrium states $\frac{\partial {\bar{W}}}{\partial x_i}|_{l,r}$ are given by
$$ \frac{\partial {\bar{W}}}{\partial x_k}|_{l} = \frac{\bar{W}-W^l}{x_k^l} , \ \ \ \ \frac{\partial {\bar{W}}}{\partial x_k}|_{r} = \frac{W^r-\bar{W}}{x_k^r},\quad (k = 1, \cdots, 3),$$
where the $x_k^{l,r}$ denote the left and right cell centers around the cell interface.

Up to now, we have presented the gas kinetic scheme (GKS) for the equations (\ref{AP3.1}). Then, after updating the flow variables inside each cell,
the radiation equations (\ref{Ap3.3}) will be solved next.

\subsection{Unified gas-kinetic scheme for radiative transfer}

\subsubsection{General formulation}

After advancing the fluid variables $(\rho,\rho\vec{v},\rho E)$ by GKS from time step $t^n$ to $t^{n+1}$, the fluid density is updated from $\rho^{n}$ to $\rho^{n+1}$,
but the fluid velocity is updated from $\vec{v}^n$ to the intermediate state $\vec{v}^h$, the same as the total specific energy from $E^n$ to $E^h$.
Therefore, the intermediate specific internal energy and kinetic energy get to $e^h$ and $\frac{1}{2}|\vec{v}^h|^2$, respectively.
Based on the updated flow values $(\rho^{n+1}, \vec{v}^h, E^h)$, the radiative transfer equations (\ref{Ap3.3}) become,
\begin{equation}
\left\{\be
\partial_t(\rho\vec{v})  = -\frac{1}{c}\displaystyle{\int\vec{\Omega} S d\vec{\Omega}}, \\[2mm]
\partial_t(\rho E) = -\frac{1}{\epsilon}\displaystyle{\int S d\vec{\Omega}}, \\[2mm]
\displaystyle{\frac{\epsilon}{c}\frac{\partial I}{\partial t} }
+\vec{\Omega}\cdot \nabla I+\epsilon \nabla \cdot (\theta \vec{\beta} I) = S. \ee \right.\label{AP3.9}
\end{equation}
For the radiation intensity in the above equations, the discrete ordinate method is used to discretize the angular variable $\vec{\Omega}$.
The vector $\vec{\Omega}$ in unit sphere is divided into $M$ discrete directions $\vec{\Omega}_m$ with corresponding integration weight $\omega_m$.
Then, the above system (\ref{AP3.9}) can be rewritten (in discrete directions) as
\begin{equation}
\left\{\be
\partial_t(\rho\vec{v})  = -\frac{1}{c}\displaystyle{\sum_{m=1}^M \vec{\Omega}_m S_m \omega_m}, \\[2mm]
\partial_t(\rho E) = -\frac{1}{\epsilon}\displaystyle{\sum_{m=1}^M S_m \omega_m}, \\[2mm]
\displaystyle{\frac{\epsilon}{c}\frac{\partial I_m}{\partial t} }
+\vec{\Omega}_m\cdot \nabla I_m+\epsilon \nabla \cdot (\theta \vec{\beta} I_m) = S_m, \quad m=1,\cdots, M,  \ee \right.\label{AP3.10}
\end{equation}
where $S_m$ is the discrete angle value of $S$ from the intensity $I$ at the discrete angle value $I_m$.

The above equations will be solved by UGKS \cite{SJXu}.
In the 2D case, the computational cells are denoted by $\{(x,y):[x_{i-\frac{1}{2}}, x_{i+\frac{1}{2}}]\times[y_{j-\frac{1}{2}},y_{j-\frac{1}{2}}]\}$.
The discrete conservation laws for the control volume  $[x_{i-\frac{1}{2}}, x_{i+\frac{1}{2}}]\times[y_{j-\frac{1}{2}},y_{j-\frac{1}{2}}]$
over the time interval $[t^n, t^{n+1}]$ for every $\vec{\Omega}_m=(\mu_m, \xi_m)$ ($m=1,\cdots, M$) are
\begin{equation}
\left\{\be
\rho^{n+1}_{i,j}(\vec{v}^{n+1}_{i,j}- \vec{v}_{i,j}^h)  = -\frac{\Delta t}{c}\displaystyle{\sum_{m=1}^M \vec{\Omega}_m S^{n+1}_{i,j,m} \omega_m}, \\[2mm]
\rho^{n+1}_{i,j} ( E^{n+1}_{i,j}-E^h_{i,j}) = -\frac{\Delta t}{\epsilon}\displaystyle{\sum_{m=1}^M S^{n+1}_{i,j,m} \omega_m}, \\[2mm]
\displaystyle{\frac{\epsilon}{c}\frac{I^{n+1}_{i,j,m}-I^{n}_{i,j,m}}{\Delta t} }
+\frac{F_{i+\frac{1}{2},j,m}-F_{i-\frac{1}{2},j,m}}{\Delta x_i\Delta y_j}+\frac{G_{i,j+\frac{1}{2},m}-G_{i,j-\frac{1}{2},m}}{\Delta x_i\Delta y_j} = S_{i,j,m}^{n+1}.  \ee \right.\label{AP3.11}
\end{equation}
Here $\Delta t = t^{n+1}-t^n, \Delta x_i = x_{i+\frac{1}{2}}-x_{i-\frac{1}{2}}$ and $\Delta y_j = y_{j+\frac{1}{2}}-y_{j-\frac{1}{2}}$. The boundary fluxes are given by
\begin{equation}
\be
F_{i+\frac{1}{2},j,m} = \int_{t^n}^{t^{n+1}}\int_{y_{j-\frac{1}{2}}}^{y_{j+\frac{1}{2}}}\mu_m I_{i+\frac{1}{2},j,m}dydt+\int_{t^n}^{t^{n+1}}\int_{y_{j-\frac{1}{2}}}^{y_{j+\frac{1}{2}}}\epsilon \theta \tilde{\beta}_x\tilde{I}_{i+\frac{1}{2},j,m}dydt \\
\qquad\qquad \triangleq F_{i+\frac{1}{2},j,m}^1+F_{i+\frac{1}{2},j,m}^2 , \\
F_{i-\frac{1}{2},j,m} = \int_{t^n}^{t^{n+1}}\int_{y_{j-\frac{1}{2}}}^{y_{j+\frac{1}{2}}}\mu_m I_{i-\frac{1}{2},j,m}dydt+\int_{t^n}^{t^{n+1}}\int_{y_{j-\frac{1}{2}}}^{y_{j+\frac{1}{2}}}\epsilon \theta \tilde{\beta}_x\tilde{I}_{i-\frac{1}{2},j,m}dydt\\
\qquad\qquad\triangleq F_{i-\frac{1}{2},j,m}^1+F_{i-\frac{1}{2},j,m}^2, \\
G_{i,j+\frac{1}{2},m} = \int_{t^n}^{t^{n+1}}\int_{x_{i-\frac{1}{2}}}^{x_{i+\frac{1}{2}}}\xi_m I_{i,j+\frac{1}{2},m}dxdt+\int_{t^n}^{t^{n+1}}\int_{x_{i-\frac{1}{2}}}^{x_{i+\frac{1}{2}}}\epsilon \theta \tilde{\beta}_y\tilde{I}_{i,j+\frac{1}{2},m}dxdt\\
\qquad\qquad \triangleq G_{i,j+\frac{1}{2},m}^1+G_{i,j+\frac{1}{2},m}^2, \\
G_{i,j-\frac{1}{2},m} = \int_{t^n}^{t^{n+1}}\int_{x_{i-\frac{1}{2}}}^{x_{i+\frac{1}{2}}}\xi_m I_{i,j-\frac{1}{2},m}dxdt+\int_{t^n}^{t^{n+1}}\int_{x_{i-\frac{1}{2}}}^{x_{i+\frac{1}{2}}}\epsilon \theta \tilde{\beta}_y\tilde{I}_{i,j-\frac{1}{2},m}dxdt \\
\qquad\qquad \triangleq G_{i,j-\frac{1}{2},m}^1+G_{i,j-\frac{1}{2},m}^2, \\
S_{i,j,m}^{n+1} = - (\frac{\sigma_t}{\epsilon})_{i,j}^{n+1}I_{i,j,m}^{n+1} + ( \frac{\sigma_t}{\epsilon}-\epsilon \sigma_s)_{i,j}^{n+1} \frac{1}{2\pi}ac(T_{i,j}^{n+1})^4 + (\frac{\epsilon \sigma_s }{2\pi})_{i,j}^{n+1}c(E_r)_{i,j}^{n+1}\\[2mm]
\qquad\qquad-\frac{1}{2\pi}(\sigma_t)_{i,j}^{n+1}\vec{\beta}_{i,j}^a\cdot [(\vec{F}_r)_{i,j}^{n+1}-(\frac{4}{3}-\theta)\epsilon (E_r)_{i,j}^{n+1}\vec{\tilde{v}}_{i,j}  ]\\[2mm]
\qquad\qquad+\frac{3}{2\pi}(\frac{4}{3}-\theta)(\sigma_t)_{i,j}^{n+1}(E_r)_{i,j}^{n+1}\vec{\Omega}\cdot\vec{\tilde{v}}_{i,j}.
\ee \label{AP3.12}
\end{equation}
With $\vec{\beta}=(\beta_x,\beta_y)$ and $\vec{\beta}_{i,j}^a=(\vec{\beta}_{i,j}^{n+1}+\vec{\beta}_{i,j}^h)/2$, $\vec{\tilde{v}}_{i,j}=\vec{v}_{i,j}^h$,
the conservation of total momentum and total energy in (\ref{Ap2.2}) can be kept. The boundary values of $\tilde{I}$ in $F_{i\pm\frac{1}{2},j,m}^2$ and $G_{i,j\pm\frac{1}{2},m}^2$
 are obtained explicitly through the upwinding according to the fluid velocity $\vec{\tilde{v}}$ on the boundary.
 In order to solve Eqs. (\ref{AP3.11}) completely, two key points have to be clarified. One is the determination of the boundary intensity $I$ in (\ref{AP3.12}) in order
 to evaluate the numerical boundary fluxes $F_{i\pm\frac{1}{2},j,m}^1$ and $G_{i,j\pm\frac{1}{2},m}^1$. Another one is to get the macroscopic variables $T, E_r$ and $\vec{F}_r$
 at time step $t^{n+1}$ in order to discretize the source term $S_{i,j,m}^{n+1}$ implicitly.

For the cell interface radiation intensity, we now give the solution in the integral form of the radiative transfer equations at the boundary.
Denote $\phi=acT^4$, around the center of a cell interface $\vec{x}_s=(x_{i-\frac{1}{2}}, y_j)$, the radiative transfer equation becomes
\begin{eqnarray}\left\{\be
\frac{\epsilon}{c}\partial_t I_{m} + \mu_m
\partial_{x}I_{m} +\epsilon \partial_x(\theta\tilde{\beta}_x \tilde{I}_m)= {(\frac{\sigma_t}{\epsilon}-\epsilon\sigma_s)}\frac{\tilde{\phi}}{2\pi}+ {\epsilon\sigma_s}\frac{c\tilde{E}_r}{2\pi} -\frac{\sigma_t}{\epsilon}I_{m} +\bar{S}_{m}, \\
\bar{S}_{m} = -\frac{1}{2\pi}\sigma_t\vec{\beta}^{a}\cdot [\vec{F}_r-(\frac{4}{3}-\theta)\epsilon \bar{E}_r\vec{\tilde{v}}  ]+\frac{3}{2\pi}(\frac{4}{3}-\theta)\sigma_t\bar{E}_r\vec{\Omega}\cdot\vec{\tilde{v}}, \\
I_{m}(x,y_j,t)|_{t=t^n} = I_{m,0}(x,y_j).
\ee \right.
  \label{3.5}
\end{eqnarray}
Here we should remark that the initial intensity $I_{m,0}$ and the functions $\tilde{\phi}, \tilde{E}_r, \vec{F}_r$ will be determined later.

Solving the above equation, the solution of (\ref{3.5}) can be represented by
\begin{eqnarray}  \be
I_{m}(t,x_{i-1/2},y_j,\mu_m,\xi_m) =\displaystyle{ \int_{t^n}^{t}\frac{c}{\epsilon}e^{-\sigma_{i-\frac{1}{2},j}(t-s)}(\bar{S}_m -\epsilon \partial_x(\theta\tilde{\beta}_x \tilde{I}_m)) ds} \\[2mm]
 \ \ \ \ \ \ + \displaystyle{ e^{-\sigma_{i-1/2,j}(t-t^n)}I_{m,0}\Big( x_{i-1/2}-\frac{c\mu_m}{\epsilon}(t-t^n)\Big) }  \\[2mm]
 \ \ \ \ \ \ + \displaystyle{ \int_{t^n}^{t}\frac{c}{\epsilon}e^{-\sigma_{i-\frac{1}{2},j}(t-s)}\Big( (\frac{\sigma_t}{\epsilon}-\epsilon\sigma_s)\frac{\tilde{\phi}}{2\pi}+
{\epsilon\sigma_s}\frac{c\tilde{E}_r}{2\pi}\Big)\big( s,x_{i-1/2} -\frac{c\mu_m}{\epsilon}(t-s)\big)ds,}
\ee \label{3.6}
\end{eqnarray}
where $\sigma = \frac{c\sigma_t}{\epsilon^2} $ and $\sigma_{i-1/2,j}$ is the value of $\sigma$ at the corresponding cell interface. Moreover,
in order to keep the asymptotic preserving property in the scheme, the value $\bar{E}_r$ in $\bar{S}_m$ should be consistently determined from $\tilde{I}$ in the boundary flux $F_{i-\frac{1}{2},j,m}^2$,
 such as
 $$\bar{E}_r=\int \tilde{I}d\vec{\Omega}=\sum^{i=M}_{i=1}\tilde{I}_m\omega_m , \quad \vec{\tilde{v}}=\vec{v}^h\;\mbox{ and }\; \vec{\beta}^a=\frac{1}{2}(\vec{v}^h+\vec{v}^{n+1}). $$
  The derivative term $\epsilon \partial_x(\theta\tilde{\beta}_x \tilde{I}_m)$ is given by
$$\epsilon \partial_x(\theta\tilde{\beta}_x \tilde{I}_m) = \frac{ 2\epsilon\theta( (\tilde{\beta}_x \tilde{I}_m)|_{(i,j)} - (\tilde{\beta}_x \tilde{I}_m)|_{(i-1,j)})}{(\Delta x_{i}+\Delta x_{i-1})}. $$

In order to determine the boundary intensity $I$ in (\ref{3.6}) completely,
the initial data $I_{m,0}$ is reconstructed by a  piecewise polynomial
\begin{equation}I_{m,0}(x,y_j) = \left\{ \begin{array}{ll} I^{n}_{i-1,j,m}+ \delta_x I_{i-1,j,m}^{n} ( x-x_{i-1,j} ), \ \ \ \ \ \mbox{ if $x<x_{i-1/2,j}$,} \\[1mm]
I^{n}_{i,j,m}+ \delta_x I_{i,j}^{n} ( x-x_{i,j} ), \ \ \ \ \ \ \
\mbox{ if $x>x_{i-1/2,j}$.} \end{array} \right. \label{3.7.1}
\end{equation}
The two spatial derivatives $\delta_x I_{i,j}^{n}$ and $\delta_x I_{i-1,j,m}^{n}$ are the reconstructed slopes at cell center $(i,j)$ and $(i-1,j)$ in the $x$-direction, respectively.
In order to remove possible numerical oscillations, the second order MUSCL-type limiter \cite{vL} is used in (\ref{3.7.1}).

The quantities $\tilde{\phi}$ and $\tilde{E}_r$ are reconstructed implicitly in time by piecewise polynomials. For the variable $\tilde{\phi}$, the reconstruction read as
\begin{eqnarray}  \be
\tilde{\phi}(x,y_j,t) = \phi_{i-1/2,j}^{n+1} + \delta_t\phi_{i-1/2,j}^{n+1}(t-t^{n+1})  \\
 \qquad + \left\{ \begin{array}{ll}  \delta_x\phi_{i-1/2,j}^{n+1,L}(x-x_{i-1/2,j}),  \mbox{ if $x<x_{i-1/2,j}$,}  \\
 \delta_x\phi_{i-1/2,j}^{n+1,R}(x-x_{i-1/2,j}),  \mbox{ if $x>x_{i-1/2,j}$, } \\
 \end{array}  \right.  \ee \label{n1.10}
\end{eqnarray}
where $\delta_t\phi_{i-1/2,j}^{n+1}=(\phi^{n+1}_{i-1/2,j}-\phi^{n}_{i-1/2,j})/\Delta t$ is the time derivative, and the spatial derivatives are
$$\delta_x\phi_{i-1/2,j}^{n+1,L}= \frac{\phi_{i-1/2,j}^{n+1}-\phi_{i-1,j}^{n+1}}{\Delta x_{i-1}/2}, \ \
 \delta_x\phi_{i-1/2,j}^{n+1,R}= \frac{\phi_{i,j}^{n+1}-\phi_{i-1/2,j}^{n+1}}{\Delta x_{i}/2}.$$
 The reconstruction for $\tilde{E}_r$ can be done in the same manner.

Finally, we turn to deal with the term $\bar{S}_m$ in the representation (\ref{3.6}). In order to keep the asymptotic preserving property of the scheme,
this term should be given consistently with the terms $F_{i\pm\frac{1}{2},j,m}^2$ and $G_{i,j\pm\frac{1}{2},m}^2$ in (\ref{AP3.12}), where the upwind side cell center value
is used by the sign of the fluid velocity $\vec{\tilde{v}}$ at the boundary.

Up to now, the formulation of the evaluation of the cell interface radiation intensity $I$ for flux evaluation has been given,
but its final determination depends on the solution of the macroscopic variables
$\phi_{i-1/2,j}^{n+1}$, $\phi_{i-1,j}^{n+1}$, $\phi_{i,j}^{n+1}$ and $(E_r)_{i-1/2,j}^{n+1}$, $(E_r)_{i-1,j}^{n+1}$, $(E_r)_{i,j}^{n+1}$.
These unknowns in $I$ and the other unknowns in the source term $S_{i,j,m}^{n+1}$ of (\ref{AP3.12}) will be determined in the next subsection.

\subsubsection{Evaluation of the macroscopic variables}

In this subsection we shall determine the macroscopic variables in the boundary fluxes and source term.
Instead of solving the radiative transfer equations, we first get the radiation energy and its transport equations
by taking moments of the third equation in (\ref{Ap2.1}), and then solve them together with the fluid dynamics equations:
\begin{equation}
\left\{\be
\partial_t(\rho\vec{v})  = -\frac{1}{c}\{-\frac{\sigma_t}{\epsilon}\vec{F}_r +(\frac{4}{3}-\theta)\sigma_tE_r\vec{{v}}\}, \\[2mm]
\partial_t(\rho E) = -\frac{1}{\epsilon}\{( \frac{\sigma_t}{\epsilon}-\epsilon \sigma_s)(acT^4-cE_r)-\sigma_t\vec{\beta}\cdot [\vec{F}_r-(\frac{4}{3}-\theta)\epsilon E_r\vec{{v}}]\}, \\[2mm]
\displaystyle{\epsilon\frac{\partial E_r}{\partial t} }
+<\vec{\Omega}\cdot \nabla I>+\epsilon \nabla \cdot <\theta \vec{{\beta}} I> = ( \frac{\sigma_t}{\epsilon}-\epsilon \sigma_s)(acT^4-cE_r) \\[2mm]
\qquad\qquad-\sigma_t\vec{\beta}\cdot [\vec{F}_r-(\frac{4}{3}-\theta)\epsilon E_r\vec{{v}}], \\[2mm]
\displaystyle{\frac{\epsilon}{c^2}\frac{\partial \vec{F}_r}{\partial t} }
+\frac{1}{c}<\vec{\Omega}\otimes\vec{\Omega}\cdot \nabla I>+\frac{\epsilon}{c} \nabla \cdot <\theta \vec{{\beta}} \otimes \vec{\Omega} I> = -\frac{\sigma_t}{c\epsilon}\vec{F}_r  \\[2mm]
\qquad\qquad+\frac{1}{c}(\frac{4}{3}-\theta)\sigma_tE_r\vec{{v}},\ee \right.
\label{n2.4}
\end{equation}
 where the angular integrations are
 $$\be
 {<\vec{\Omega} \cdot \nabla I>}:=\int\vec{\Omega} \nabla I d\vec{\Omega}, \ \ <\vec{\Omega}\otimes\vec{\Omega}\cdot \nabla I>:=\int\vec{\Omega}\otimes \vec{\Omega} \nabla I d\vec{\Omega};\\
  {<\theta \vec{\beta} I>}:=\int\theta \vec{\beta} I d\vec{\Omega}, \ \ <\theta \vec{{\beta}}\otimes \vec{\Omega} I>:=\int\theta\vec{\beta}\otimes \vec{\Omega} I d\vec{\Omega}.
 \ee$$

The finite volume method for the system (\ref{n2.4}) reads as follows.
 \begin{equation}
\left\{\be
\rho_{i,j}^{n+1}\vec{v}_{i,j}^{n+1}  = \rho_{i,j}^{n+1}\vec{v}_{i,j}^{h} -\frac{\Delta t}{c}\{-\frac{\sigma_{t,i,j}^{n+1}}{\epsilon}(\vec{F}_r)_{i,j}^{n+1} +(\frac{4}{3}-\theta)\sigma_{t,i,j}^{n+1}(E_r)_{i,j}^{n+1}\vec{{v}}_{i,j}^h\}, \\[2mm]
\rho_{i,j}^{n+1} E_{i,j}^{n+1} = \rho_{i,j}^{n+1} E_{i,j}^{h}-\frac{\Delta t}{\epsilon}\{( \frac{\sigma_{t,i,j}^{n+1}}{\epsilon}-\epsilon \sigma_{s,i,j}^{n+1})(ac(T_{i,j}^{n+1})^4-c(E_r)_{i,j}^{n+1})-\\[2mm]
\qquad\qquad \sigma_{t,i,j}^{n+1}\vec{\beta}_{i,j}^a\cdot [(\vec{F}_r)_{i,j}^{n+1}-(\frac{4}{3}-\theta)\epsilon (E_r)_{i,j}^{n+1}\vec{{v}}_{i,j}^h]\}, \\[2mm]
\displaystyle{\epsilon (E_r)_{i,j}^{n+1} }
+\frac{\Delta t}{\Delta x_i\Delta y_j}(\Phi_{i+\frac{1}{2},j}^{n+1} - \Phi_{i-\frac{1}{2},j}^{n+1} )+\frac{\Delta t}{\Delta x_i\Delta y_j}(\Psi_{i,j+\frac{1}{2}}^{n+1} - \Psi_{i,j-\frac{1}{2}}^{n+1} )= \\[2mm] \qquad\qquad\epsilon (E_r)_{i,j}^{n} + \Delta t\{ ( \frac{\sigma_{t,i,j}^{n+1}}{\epsilon}-\epsilon \sigma_{s,i,j}^{n+1})(ac(T_{i,j}^{n+1})^4-c(E_r)_{i,j}^{n+1})- \\[2mm] \qquad\qquad\sigma_{t,i,j}^{n+1}\vec{\beta}_{i,j}^a\cdot [(\vec{F}_r)_{i,j}^{n+1}-(\frac{4}{3}-\theta)\epsilon (E_r)_{i,j}^{n+1}\vec{{v}}_{i,j}^h]\}, \\[2mm]
\displaystyle{\frac{\epsilon}{c^2} (\vec{F}_r)_{i,j}^{n+1} }
+\frac{\Delta t}{\Delta x_i\Delta y_j}(\vec{\bar{\Phi}}_{i+\frac{1}{2},j}^{n+1} - \vec{\bar{\Phi}}_{i-\frac{1}{2},j}^{n+1} )+\frac{\Delta t}{\Delta x_i\Delta y_j}(\vec{\bar{\Psi}}_{i,j+\frac{1}{2}}^{n+1} - \vec{\bar{\Psi}}_{i,j-\frac{1}{2}}^{n+1} ) =  \\[2mm]
\qquad\qquad \displaystyle{\frac{\epsilon}{c^2} (\vec{F}_r)_{i,j}^{n} } + \Delta t\{ -\frac{\sigma_{t,i,j}^{n+1}}{c\epsilon}(\vec{F}_r)_{i,j}^{n+1} +\frac{1}{c}(\frac{4}{3}-\theta)\sigma_{t,i,j}^{n+1}(E_r)_{i,j}^{n+1}\vec{{v}}_{i,j}^h\},\ee \right.
\label{n2.5}
\end{equation}
where ${\vec{\beta}}_{i,j}^a=({\vec{\beta}}_{i,j}^{n+1}+{\vec{\beta}}_{i,j}^h)/2$.

It should be emphasized that the central ingredient in UGKS is about the use of the same time evolution distribution function for the microscopic and macroscopic fluxes at a cell interface \cite{xu2010unified}.
To be consistent with this methodology, the boundary fluxes in (\ref{n2.5}) are obtained by angular integration of $F$ and $G$ in (\ref{AP3.12}),
 \begin{equation}\be
\Phi_{i+\frac{1}{2},j}^{n+1} = \sum_{m=1}^M F_{i+\frac{1}{2},j,m}\omega_m = \sum_{m=1}^M (F_{i+\frac{1}{2},j,m}^1+F_{i+\frac{1}{2},j,m}^2)\omega_m, \\
\Phi_{i-\frac{1}{2},j}^{n+1} = \sum_{m=1}^M F_{i-\frac{1}{2},j,m}\omega_m = \sum_{m=1}^M (F_{i-\frac{1}{2},j,m}^1+F_{i-\frac{1}{2},j,m}^2)\omega_m, \\
\Psi_{i,j+\frac{1}{2}}^{n+1} = \sum_{m=1}^M G_{i,j+\frac{1}{2},m}\omega_m = \sum_{m=1}^M (G_{i,j+\frac{1}{2},m}^1+G_{i,j+\frac{1}{2},m}^2)\omega_m, \\
\Psi_{i,j-\frac{1}{2}}^{n+1} = \sum_{m=1}^M G_{i,j-\frac{1}{2},m}\omega_m = \sum_{m=1}^M (G_{i,j-\frac{1}{2},m}^1+G_{i,j-\frac{1}{2},m}^2)\omega_m, \\
\vec{\bar{\Phi}}_{i+\frac{1}{2},j}^{n+1} =\frac{1}{c} \sum_{m=1}^M\vec{\Omega}_m F_{i+\frac{1}{2},j,m}\omega_m = \frac{1}{c}\sum_{m=1}^M \vec{\Omega}_m (F_{i+\frac{1}{2},j,m}^1+F_{i+\frac{1}{2},j,m}^2)\omega_m, \\
\vec{\bar{\Phi}}_{i-\frac{1}{2},j}^{n+1} = \frac{1}{c}\sum_{m=1}^M \vec{\Omega}_m F_{i-\frac{1}{2},j,m}\omega_m = \frac{1}{c}\sum_{m=1}^M \vec{\Omega}_m (F_{i-\frac{1}{2},j,m}^1+F_{i-\frac{1}{2},j,m}^2)\omega_m, \\
\vec{\bar{\Psi}}_{i,j+\frac{1}{2}}^{n+1} =\frac{1}{c} \sum_{m=1}^M \vec{\Omega}_m G_{i,j+\frac{1}{2},m}\omega_m = \frac{1}{c}\sum_{m=1}^M \vec{\Omega}_m (G_{i,j+\frac{1}{2},m}^1+G_{i,j+\frac{1}{2},m}^2)\omega_m, \\
\vec{\bar{\Psi}}_{i,j-\frac{1}{2}}^{n+1} = \frac{1}{c}\sum_{m=1}^M \vec{\Omega}_m G_{i,j-\frac{1}{2},m}\omega_m = \frac{1}{c}\sum_{m=1}^M \vec{\Omega}_m (G_{i,j-\frac{1}{2},m}^1+G_{i,j-\frac{1}{2},m}^2)\omega_m.
\ee\label{n2.6}
\end{equation}
Thus, based on the macroscopic interface fluxes in (\ref{n2.6}), the system (\ref{n2.5}) reduces to a coupled nonlinear system of the macroscopic quantities $\vec{v}_{i,j}^{n+1}, T_{i,j}^{n+1}, (E_r)_{i,j}^{n+1}$
and $(\vec{F}_r)_{i,j}^{n+1}$ only, where the parameters $\sigma_{t,i,j}^{n+1}$ and $\sigma_{s,i,j}^{n+1}$ depend implicitly on the material temperature $T_{i,j}^{n+1}$.
This nonlinear system can be solved by iterative method, such as the Gauss-Seidel iteration method as shown in \cite{SJXu, SJX2}.

\subsubsection{Update of the solution}

After obtaining the macroscopic variables $T_{i,j}^{n+1}, (E_r)_{i,j}^{n+1}$ and $(\vec{F}_r)_{i,j}^{n+1}$ by solving the
equations (\ref{n2.5}) iteratively, we can fully determine the radiation intensity at the cell interface for the microscopic flux evaluation.
For example, the boundary value ${\phi}_{i-\frac{1}{2},j}^{n+1}$ in (\ref{n1.10}) is given by
$$ \phi_{i,j-1/2}^{k+1}=(\phi_{i,j}^{k+1}+\phi_{i,j-1}^{k+1})/2 .$$
The left and right derivatives in (\ref{n1.10}) are given by
$$\delta_x\phi_{i-1/2,j}^{n+1,L}= \frac{\phi_{i-1/2,j}^{n+1}-\phi_{i-1,j}^{n+1}}{\Delta x_{i-1}/2}, \ \ \ \ \
 \delta_x\phi_{i,j-1/2}^{n+1,R}= \frac{\phi_{i,j}^{n+1}-\phi_{i-1/2,j}^{n+1}}{\Delta x_i/2}. $$
For the time derivative $\delta_t \phi_{i-1/2,j}^{n+1}$ in (\ref{n1.10}), we can take
$$\delta_t \phi_{i-1/2,j}^{n+1} = \frac{ \phi_{i-1/2,j}^{n+1} - \phi_{i-1/2,j}^{n} }{\Delta t}.$$
In the same way, the reconstruction of $\tilde{E}_r$ in (\ref{3.6}) can be obtained.

With the determined macroscopic variables in (\ref{n2.5}), the source term $S_{i,j,m}^{n+1}$ and the numerical boundary fluxes $F_{i\pm\frac{1}{2},j,m}, G_{i,j\pm\frac{1}{2},m}$ in (\ref{AP3.12})
can be then explicitly evaluated. Afterwards, the radiative intensity in (\ref{AP3.11}) can be updated as follows.
\begin{equation}\left\{ \be
\hat{S}_{i,j,m}^{n+1} \triangleq
( \frac{\sigma_t}{\epsilon}-\epsilon \sigma_s)_{i,j}^{n+1} \frac{1}{2\pi}ac(T_{i,j}^{n+1})^4 + (\frac{\epsilon \sigma_s }{2\pi})_{i,j}^{n+1}c(E_r)_{i,j}^{n+1}\\[2mm]
\qquad\qquad-\frac{1}{2\pi}(\sigma_t)_{i,j}^{n+1}\vec{\beta}_{i,j}^a\cdot [(\vec{F}_r)_{i,j}^{n+1}-(\frac{4}{3}-\theta)\epsilon (E_r)_{i,j}^{n+1}\vec{\tilde{v}}_{i,j}  ]\\[2mm]
\qquad\qquad+\frac{3}{2\pi}(\frac{4}{3}-\theta)(\sigma_t)_{i,j}^{n+1}(E_r)_{i,j}^{n+1}\vec{\Omega}\cdot\vec{\tilde{v}}_{i,j}, \\[2mm]
\displaystyle{I_{i,j}^{n+1} = \frac{\frac{\epsilon}{c\Delta t}I_{i,j}^n + \frac{F_{i-\frac{1}{2},j,m}
- F_{i+\frac{1}{2},j,m}}{\Delta x_i \Delta y_j}+ \frac{G_{i,j-\frac{1}{2},m} - G_{i,j+\frac{1}{2}},m}{\Delta x_i \Delta y_j}+\hat{S}_{i,j,m}^{n+1}}{\frac{\epsilon}{c\Delta t}+(\frac{\sigma_t}{c})_{i,j}^{n+1}}}.
 \ee \right. \label{4.5}
\end{equation}
This completes the main numerical procedures in our unified gas kinetic scheme.

The final step is to update the solutions in the first and second equations in (\ref{AP3.11}) for the fluid velocity $\vec{v}_{i,j}^{n+1}$
and material temperature $T_{i,j}^{n+1}$
with the newly obtained value $I_{i,j,m}^{n+1}$.
The solutions for the momentum equations (\ref{AP3.11}) and the energy equation (\ref{AP3.11}) are given respectively by
\begin{equation}\left\{
\be
\displaystyle{\vec{v}_{i,j}^{n+1} = \frac{\rho_{i,j}^{n+1}\vec{v}_{i,j}^h - \frac{\Delta t}{c}\sum_{m=1}^M\vec{\Omega}_m(\hat{S}_{i,j,m}^{n+1}+ (\frac{\sigma_t}{c})_{i,j}^{n+1}I_{i,j,m}^{n+1})\omega_m }{\rho_{i,j}^{n+1}}}, \\[2mm]
\displaystyle{E_{i,j}^{n+1} = \frac{\rho_{i,j}^{n+1}E_{i,j}^h - \frac{\Delta t}{\epsilon}\sum_{m=1}^M(\hat{S}_{i,j,m}^{n+1} + (\frac{\sigma_t}{c})_{i,j}^{n+1}I_{i,j,m}^{n+1})\omega_m}{\rho_{i,j}^{n+1}}}, \\[2mm]
\displaystyle{T_{i,j}^{n+1} = \frac{E_{i,j}^{n+1} - |\vec{v}_{i,j}^{n+1}|^2/2}{C_v}}.
 \ee \right. \label{4.6}
\end{equation}
Based on (\ref{4.5}) and (\ref{4.6}), we get the solution of the system (\ref{AP3.9}), and complete the construction of our GKS and UGKS for the radiation hydrodynamic system.
In the following section, the asymptotic preserving property of the proposed scheme will be analyzed.

\setcounter{equation}{0}
\section{Asymptotic analysis of the scheme}\label{sec:APanal}

The scheme presented in the last section possesses the asymptotic preserving (AP) property. In fact, following the analysis in \cite{SJXu, SJX2},
we are able to show such a property for the scheme in capturing the diffusion solution in the optically thick region for the radiative transfer.
The numerical fluxes $F$ and $G$ in (\ref{AP3.12}) play a dominant role in the proof of the AP property.
Firstly, the left boundary numerical flux in the $x$-direction is given by
$$ \frac{c}{\epsilon}F_{i,j-1/2,m}^1 = \frac{ c\mu_{m}\Delta y_j}{\epsilon\Delta t}\int_{t^n}^{t^{n+1}}I_{m}(t,x_{i-1/2},y_{j},\mu_{m},\xi_{m})dt, $$
which can be exactly evaluated as follows. Using (\ref{3.6}), we get from the above identity that
\begin{equation} \be
\frac{c}{\epsilon}F_{i-1/2,j,m}^1 = \Delta y_j\{A_{i-1/2,j}\mu_{m}(
I_{i-1/2,j,m}^{n,-}\mathrm{1}_{\mu_{m}>0}+I_{i-1/2,j,m}^{n,+}\mathrm{1}_{\mu_{m}<0}) \\
\ \ \ \ \ \ \ \ \ \ \ \ \ + D_{i-1/2,j}^1( \mu_{m}^2
\delta_x\phi_{i-1/2,j}^{n+1,L}\mathrm{1}_{\mu_{m}>0}+
\mu_{m}^2\delta_x\phi_{i-1/2,j}^{n+1,R}\mathrm{1}_{\mu_{m}<0})\\
\ \ \ \ \ \ \ \ \ \ \ \ \ + D_{i-1/2,j}^2( \mu_{m}^2
\delta_x\varphi_{i-1/2,j}^{n+1,L}\mathrm{1}_{\mu_{m}>0}+
\mu_{m}^2\delta_x\varphi_{i-1/2,j}^{n+1,R}\mathrm{1}_{\mu_{m}<0}) \\
 \ \ \ \ \ \ \ \ \ \ \ \ \ + B_{i-1/2,j}( \mu_{m}^2
\delta_xI_{i-1,j,m}^{n}\mathrm{1}_{\mu_{m}>0}+
\mu_{m}^2\delta_xI_{i,j,m}^{n}\mathrm{1}_{\mu_{m}<0}) \\
\ \ \ \ \ \ \ \ \ \ \ \ \ +
E_{i-1/2,j}^1\mu_{m}\delta_t\phi_{i-1/2,j}^{n+1} + C_{i-1/2,j}^1\mu_{m} \phi_{i-1/2,j}^{n+1}\\
\ \ \ \ \ \ \ \ \ \ \ \ \ +
E_{i-1/2,j}^2\mu_{m}\delta_t\varphi_{i-1/2,j}^{n+1} + C_{i-1/2,j}^2\mu_{m} \varphi_{i-1/2,j}^{n+1}\\
\ \ \ \ \ \ \ \ \ \ \ \ \ + P_{i-1/2,j}\mu_{m}(\bar{S}_m -\epsilon \partial_x(\theta\tilde{\beta}_x \tilde{I}_m))|_{i-\frac{1}{2},j}\}.  \ee \label{Ap2.15.1}
\end{equation}
Here $I_{i-1/2,j,m}^{n,-}, I_{i-1/2,j,m}^{n,+}$ are the interface values given by
\begin{equation} \be
I_{i-1/2,j,m}^{n,-} = I_{i-1,j,m}^n + \delta_x I_{i-1,j,m}^{n}(x_{i-1/2}-x_{i-1,j}), \\
I_{i-1/2,j,m}^{n,+} = I_{i,j,m}^n + \delta_x I_{i,j,m}^{n}(x_{i-1/2}-x_{i,j}), \ee \nonumber
\end{equation}
and $\delta_x I_{i-1,j,m}^{n}$ and $\delta_x I_{i,j,m}^{n}$ are slopes in the $x$-direction which are reconstructed in (\ref{3.7.1}).

After a straightforward calculation, the coefficients in (\ref{Ap2.15.1}) are given by
\begin{equation} \be
A = \frac{c}{\epsilon \Delta t \nu }( 1 - e^{-\nu \Delta
t}), \\
C^1 = \frac{c^2 (\frac{\sigma_t}{\epsilon}-\epsilon \sigma_s ) }{2\pi\Delta t \epsilon^2 \nu }
(\Delta t - \frac{1}{\nu}(1-e^{-\nu \Delta t })),  \\
C^2 = \frac{c^2 {\epsilon}\sigma_s }{2\pi\Delta t \epsilon^2 \nu }
(\Delta t - \frac{1}{\nu}(1-e^{-\nu \Delta t })), \\
D^1 = -\frac{c^3(\frac{\sigma_t}{\epsilon}-\epsilon \sigma_s )}{2\pi\Delta t \epsilon^3 \nu^2}( \Delta
t ( 1 + e^{-\nu \Delta t}) - \frac{2}{\nu}( 1 - e^{-\nu\Delta
t}) ), \\
D^2 = -\frac{c^3 {\epsilon}\sigma_s}{2\pi\Delta t \epsilon^3 \nu^2}( \Delta
t ( 1 + e^{-\nu \Delta t}) - \frac{2}{\nu}( 1 - e^{-\nu\Delta
t}) ), \\
B = -\frac{c^2}{\epsilon^2\nu^2\Delta t}( 1 - e^{-\nu\Delta t}
-\nu\Delta te^{-\nu\Delta t}), \\
E^1 = \frac{c^2(\frac{\sigma_t}{\epsilon}-\epsilon \sigma_s )}{2\pi\epsilon^2 \nu^3\Delta t}(1- e^{-\nu\Delta
t}-\nu\Delta t e^{-\nu\Delta t} - \frac{1}{2}(\nu\Delta t)^2),\\
E^2 = \frac{c^2 {\epsilon}\sigma_s}{2\pi\epsilon^2\nu^3\Delta t}(1- e^{-\nu\Delta
t}-\nu\Delta t e^{-\nu\Delta t} - \frac{1}{2}(\nu\Delta t)^2), \\
P = \frac{c^2 }{\Delta t \epsilon^2 \nu } (\Delta t - \frac{1}{\nu}(1-e^{-\nu \Delta t }))
 \ee \label{Ap2.14}
\end{equation}
with $\nu = \frac{c\sigma_t}{\epsilon^2}$.

The behavior of the scheme in the small-$\epsilon$ limit is completely controlled by the limits of these coefficients, as shown in
the following proposition.
\\[2mm]
{\bf Proposition 1.} {\it Let $\sigma_t$ and $\sigma_s$ be
positive. Then, as $\epsilon$ tends to zero, we have

$\bullet$ $A(\Delta t,\epsilon,\sigma, \nu)\to 0$;

$\bullet$ $B(\Delta t,\epsilon,\sigma, \nu)\to 0$;

$\bullet$ $D^1(\Delta t,\epsilon,\sigma, \nu)\to -c/(2\pi\sigma_t)$;

$\bullet$ $D^2(\Delta t,\epsilon,\sigma, \nu)\to 0$;

$\bullet$ $P(\Delta t,\epsilon,\sigma, \nu)\to c/\sigma_t$;

$\bullet$ $\frac{\epsilon}{c^2}E^1(\Delta t,\epsilon,\sigma, \nu)\to -\Delta t/(4\pi c)$;

$\bullet$ $\frac{\epsilon}{c^2}E^2(\Delta t,\epsilon,\sigma, \nu)\to 0$;

$\bullet$ $\frac{\epsilon}{c^2}C^1(\Delta t,\epsilon,\sigma, \nu)\to 1/(2\pi c)$;

$\bullet$ $\frac{\epsilon}{c^2}C^2(\Delta t,\epsilon,\sigma, \nu)\to 0$.} \vspace{2mm}

On the other hand, when taking moment of the left boundary flux $F_{i-\frac{1}{2},j,m}$ over the propagation angle $\vec{\Omega}$, we obtain
 \begin{equation}\be
\frac{c}{\epsilon}\Phi_{i-\frac{1}{2},j}^{n+1} = \frac{c}{\epsilon}\sum_{m=1}^M F_{i-\frac{1}{2},j,m}\omega_m = \frac{c}{\epsilon}\sum_{m=1}^M (F_{i-\frac{1}{2},j,m}^1+F_{i-\frac{1}{2},j,m}^2)\omega_m \\
\qquad\qquad = \Delta y_j \{ A_{i-1/2,j}\sum_{m=1}^{M}\omega_m\mu_m \Big( I_{i-1,j,m}^n 1_{\mu_m>0} +I_{i,j,m}^n1_{\mu_m<0}\Big) \\
\qquad\qquad +{{\frac{2\pi D_{i-1/2,j}^1}{3}(\frac{\phi_{i,j}^{n+1}-\phi_{i-1,j}^{n+1}}{0.5*(\Delta x_i+\Delta x_{i-1})})}} +{{\frac{2\pi D_{i-1/2,j}^2}{3}(\frac{\varphi_{i,j}^{n+1}-\varphi_{i-1,j}^{n+1}}{0.5*(\Delta x_i+\Delta x_{i-1})})}}\\
\qquad\qquad+{(\frac{4}{3}-\theta)c\bar{E}_r{\tilde{v}}_x}|_{i-\frac{1}{2},j}-\frac{c}{\sigma_t}\sum_{m=1}^M\omega_m\Omega_m(\epsilon \partial_x(\theta\tilde{\beta}_x \tilde{I}_m))|_{i-\frac{1}{2},j}\\
\qquad\qquad + B_{i-1/2,j}\sum_{m=1}^{M}\omega_m\mu^2_m (\delta_x I^n_{i-1,j,m} 1_{\mu_m>0} +\delta_x I_{i,j,m}^n 1_{\mu_m<0})+ \theta c\bar{E}_r{\tilde{v}}_x|_{i-\frac{1}{2},j} \}\\
\qquad\qquad = \Delta y_j \{ A_{i-1/2,j}\sum_{m=1}^{M}\omega_m\mu_m \Big( I_{i-1,j,m}^n 1_{\mu_m>0} +I_{i,j,m}^n1_{\mu_m<0}\Big) \\
\qquad\qquad +{{\frac{2\pi D_{i-1/2,j}^1}{3}(\frac{\phi_{i,j}^{n+1}-\phi_{i-1,j}^{n+1}}{0.5*(\Delta x_i+\Delta x_{i-1})})}} +{{\frac{2\pi D_{i-1/2,j}^2}{3}(\frac{\varphi_{i,j}^{n+1}-\varphi_{i-1,j}^{n+1}}{0.5*(\Delta x_i+\Delta x_{i-1})})}}\\
\qquad\qquad+{\frac{4}{3}c\bar{E}_r{\tilde{v}}_x}|_{i-\frac{1}{2},j}-\frac{c\epsilon}{\sigma_t}\sum_{m=1}^M\omega_m\Omega_m( \partial_x(\theta\tilde{\beta}_x \tilde{I}_m))|_{i-\frac{1}{2},j}\\
\qquad\qquad + B_{i-1/2,j}\sum_{m=1}^{M}\omega_m\mu^2_m (\delta_x I^n_{i-1,j,m} 1_{\mu_m>0} +\delta_x I_{i,j,m}^n 1_{\mu_m<0})\}
\\
\qquad\qquad\xrightarrow[\epsilon\rightarrow0]{ }{\Delta y_j}\{-\frac{c}{3\sigma_t}\frac{\phi_{i,j}^{n+1}-\phi_{i-1,j}^{n+1}}{0.5*(\Delta x_i+\Delta x_{i-1})} + {\frac{4}{3}c(\bar{E}_r{\tilde{v}}_x})|_{i-\frac{1}{2},j}\},
\ee
\label{RH4.1}
\end{equation}
and
\begin{equation}\be
\vec{\bar{\Phi}}_{i-\frac{1}{2},j}^{n+1} = \displaystyle{ \frac{1}{c}\sum_{m=1}^M \vec{\Omega}_mF_{i-\frac{1}{2},j,m}\omega_m = \frac{1}{c}\sum_{m=1}^M \vec{\Omega}_m(F_{i-\frac{1}{2},j,m}^1+F_{i-\frac{1}{2},j,m}^2)\omega_m} \\
\qquad\qquad =\displaystyle{\frac{\epsilon}{c^2}\Delta y_j\sum_{m=1}^M \vec{\Omega}_m}\{A_{i-1/2,j}\mu_{m}(
I_{i-1/2,j,m}^{n,-}\mathrm{1}_{\mu_{m}>0}+I_{i-1/2,j,m}^{n,+}\mathrm{1}_{\mu_{m}<0}) \\
\qquad\qquad \ \ + D_{i-1/2,j}^1( \mu_{m}^2
\delta_x\phi_{i-1/2,j}^{n+1,L}\mathrm{1}_{\mu_{m}>0}+
\mu_{m}^2\delta_x\phi_{i-1/2,j}^{n+1,R}\mathrm{1}_{\mu_{m}<0})\\
\qquad\qquad \ \ + D_{i-1/2,j}^2( \mu_{m}^2
\delta_x\varphi_{i-1/2,j}^{n+1,L}\mathrm{1}_{\mu_{m}>0}+
\mu_{m}^2\delta_x\varphi_{i-1/2,j}^{n+1,R}\mathrm{1}_{\mu_{m}<0}) \\
\qquad\qquad \ \ + B_{i-1/2,j}( \mu_{m}^2
\delta_xI_{i-1,j,m}^{n}\mathrm{1}_{\mu_{m}>0}+
\mu_{m}^2\delta_xI_{i,j,m}^{n}\mathrm{1}_{\mu_{m}<0}) \\
\qquad\qquad \ \ +
E_{i-1/2,j}^1\mu_{m}\delta_t\phi_{i-1/2,j}^{n+1} + C_{i-1/2,j}^1\mu_{m} \phi_{i-1/2,j}^{n+1}\\
\qquad\qquad \ \ +
E_{i-1/2,j}^2\mu_{m}\delta_t\varphi_{i-1/2,j}^{n+1} + C_{i-1/2,j}^2\mu_{m} \varphi_{i-1/2,j}^{n+1}\\
\qquad\qquad \ \ + P_{i-1/2,j}\mu_{m}(\bar{S}_m -\epsilon \partial_x(\theta\tilde{\beta}_x \tilde{I}_m))|_{i-\frac{1}{2},j}\}\omega_m \\
\qquad\qquad\xrightarrow[\epsilon\rightarrow0]{}{\Delta y_j}\displaystyle{\sum_{m=1}^M \vec{\Omega}_m\{\frac{1}{2\pi c} \mu_m \phi_{i-1/2,j}^{n+1} -\frac{\Delta t}{4\pi c}\mu_m  \delta_t\phi_{i-1/2,j}^{n+1} \}\omega_m}\\
\qquad\qquad \ \ = \frac{\Delta y_j}{3c}(\phi_{i-1/2,j}^{n+1}- \frac{\Delta t}{2}\delta_t\phi_{i-1/2,j}^{n+1})= \frac{\Delta y_j}{3c}\frac{\phi_{i-1/2,j}^{n+1}+\phi_{i-1/2,j}^{n}}{2}.
\ee
\label{RH4.1.1}
\end{equation}
With the limits in (\ref{RH4.1}) and  (\ref{RH4.1.1}), it is easy to see that the above coupled GKS and UGKS method
possesses the asymptotic preserving property, provided the following proposition holds.
\\[2mm]
\noindent{\bf Proposition 2.} {\it Let $\sigma_t$ and $\sigma_s$ be positive. Then,
as $\epsilon$ tends to zero, the numerical scheme given by coupling (\ref{AP3.6}) for the fluid part with (\ref{AP3.11}) for the radiation part
goes to the standard implicit diffusion scheme for the equilibrium diffusion limit system (\ref{Ap2.9}) of radiation hydrodynamics.  }
\vspace{1mm}

\noindent{\bf Proof.} Firstly, as $\epsilon\to 0$, we see that the term of $\epsilon^{-1}$-order in the third equation of (\ref{AP3.11})
satisfies
\begin{equation}I_{i,j,m}^{n+1}\rightarrow\frac{1}{2\pi}\phi_{i,j}^{n+1}=\frac{1}{2\pi}ac(T_{i,j}^{n+1})^4.\label{L2.1} \end{equation}
Integrating the above equation {with respect to the angular variable, we find that
\begin{equation}c(E_r)_{i,j}^{n+1}\rightarrow\phi_{i,j}^{n+1}=
a c (T_{i,j}^{n+1} )^4, \ \  (\vec{F}_r)_{i,j}^{n+1}\rightarrow 0, \ \ \bar{\bar{P}}_{i,j}^{n+1}\rightarrow\frac{1}{3}a(T_{i,j}^{n+1} )^4\bar{\bar{D}}. \label{p2.1} \end{equation}

Secondly, we integrate the flux $\frac{1}{\epsilon}F_{i-1/2,j}^{k+1}$ in the angular variable to obtain the macro flux $\frac{1}{\epsilon}\Phi_{i-1/2,j,m,n}^{k+1}$ in (\ref{RH4.1.1}).
Then, taking $\epsilon\to 0$, we utilize Proposition 1 to obtain
\begin{equation}\frac{1}{\epsilon}\Phi_{i-1/2,j}^{k+1} \rightarrow
{\Delta y_j}\Big( -\frac{1}{3\sigma_{i-1/2,j}^{n+1}}\frac{\phi_{i,j}^{n+1}
-\phi_{i-1,j}^{n+1} }{0.5*(\Delta x_i+\Delta x_{i-1})} +\frac{4}{3}(\bar{E}_r{\tilde{v}_x})|_{i-\frac{1}{2},j}\Big). \label{p2.2}
\end{equation}
 Similarly,
as $\epsilon\rightarrow 0,$ the other macro boundary interface
fluxes imply
\begin{equation}\begin{array}{ll} \frac{1}{\epsilon}\Phi_{i+1/2,j}^{k+1} \to \displaystyle{{\Delta y_j}\Big( - \frac{1}{3\sigma_{i+1/2,j}^{n+1}}\frac{\phi_{i+1,j}^{n+1}
-\phi_{i,j}^{n+1} }{ 0.5*(\Delta x_i+\Delta x_{i+1})} +\frac{4}{3}(\bar{E}_r{\tilde{v}_x})|_{i+\frac{1}{2},j}\Big),   } \\
 \frac{1}{\epsilon}\Psi_{i,j-1/2}^{n+1} \rightarrow
\displaystyle{{\Delta x_i}\Big( - \frac{1}{3\sigma_{i,j-1/2}^{n+1}}\frac{\phi_{i,j}^{n+1}
-\phi_{i,j-1}^{n+1} }{0.5*(\Delta y_j+\Delta y_{j-1})} +\frac{4}{3}(\bar{E}_r{\tilde{v}_y})|_{i,j-\frac{1}{2}}\Big),  }\\
 \frac{1}{\epsilon}\Psi_{i,j+1/2}^{n+1} \rightarrow
\displaystyle{{\Delta x_i}\Big( - \frac{1}{3\sigma_{i,j+1/2}^{n+1}}\frac{\phi_{i,j+1}^{n+1}
-\phi_{i,j}^{n+1} }{0.5*(\Delta y_j+\Delta y_{j+1})} +\frac{4}{3}(\bar{E}_r{\tilde{v}_y})|_{i,j+\frac{1}{2}}\Big).   }
\end{array}   \label{p2.3}
\end{equation}
By dividing the third equation of (\ref{AP3.11}) by $\epsilon$ and integrating the resulting equation over the angular variable, we utilize
the second equation of (\ref{AP3.11}) and (\ref{p2.3}) to deduce that as $\epsilon\rightarrow 0$,
\begin{equation}\begin{array}{ll}
\rho_{i,j}^{n+1}\frac{E_{i,j}^{n+1}-E_{i,j}^h}{\Delta t } + \frac{(E_r)_{i,j}^{n+1}-(E_r)_{i,j}^n}{\Delta t }+\frac{1}{\Delta x_i}\Big\{( - \frac{1}{3\sigma_{i+1/2,j}^{n+1}}\frac{\phi_{i+1,j}^{n+1}
-\phi_{i,j}^{n+1} }{ 0.5*(\Delta x_i+\Delta x_{i+1})} +\frac{4}{3}(\bar{E}_r{\tilde{v}_x})|_{i+\frac{1}{2},j}) -  \\
\qquad\qquad\qquad ( - \frac{1}{3\sigma_{i-1/2,j}^{n+1}}\frac{\phi_{i,j}^{n+1}
-\phi_{i-1,j}^{n+1} }{0.5*(\Delta x_i+\Delta x_{i-1})} +\frac{4}{3}(\bar{E}_r{\tilde{v}_x})|_{i-\frac{1}{2},j})\Big\} +  \\
\qquad\qquad\qquad \frac{1}{\Delta y_j}\Big\{( - \frac{1}{3\sigma_{i,j+1/2}^{n+1}}\frac{\phi_{i,j+1}^{n+1}
-\phi_{i,j}^{n+1} }{0.5*(\Delta y_j+\Delta y_{j+1})} +\frac{4}{3}(\bar{E}_r{\tilde{v}_y})|_{i,j+\frac{1}{2}})- \\
\qquad\qquad\qquad ( - \frac{1}{3\sigma_{i,j-1/2}^{n+1}}\frac{\phi_{i,j}^{n+1}
-\phi_{i,j-1}^{n+1} }{0.5*(\Delta y_j+\Delta y_{j-1})} +\frac{4}{3}(\bar{E}_r{\tilde{v}_y})|_{i,j-\frac{1}{2}})\Big\} = 0.
\end{array}
\label{p2.3.1}
\end{equation}
Based on this equation (\ref{p2.3.1}), together with the discretization of the third equation of (\ref{AP3.1}) in the fluid part,
we get the numerical discretization of the third equation of the equilibrium diffusive radiative hydrodynamics (\ref{Ap2.9}).

Thirdly, multiplying the third equation of (\ref{AP3.11}) with $\vec{\omega}$ and integrating the resulting equation in the angular variable,
we use the first equation of (\ref{AP3.11}) and (\ref{RH4.1.1}) to infer that
\begin{equation}
\begin{array}{ll}
\rho_{i,j}^{n+1}\frac{\vec{v}_{i,j}^{n+1}-\vec{v}_{i,j}^h}{\Delta t} + \frac{1}{\Delta x_i}\{\frac{\phi_{i+1/2,j}^{n+1}+\phi_{i+1/2,j}^{n}}{6c} - \frac{\phi_{i-1/2,j}^{n+1}+\phi_{i-1/2,j}^{n}}{6c} \} + \\
\qquad\qquad\qquad\quad\frac{1}{\Delta y_j}\{\frac{\phi_{i,j+1/2}^{n+1}+\phi_{i,j+1/2}^{n}}{6c} - \frac{\phi_{i,j-1/2}^{n+1}+\phi_{i,j-1/2}^{n}}{6c} \} = 0.
\end{array}  \label{p2.4}
\end{equation}
Thus, with the discretization of the second equation of (\ref{AP3.1}) in fluid part and the above equation (\ref{p2.4}), we get  the numerical discretization of the second equation of the equilibrium diffusive radiative hydrodynamics (\ref{Ap2.9}).

Since there is no $\epsilon$ term in the first equation of (\ref{Ap2.1}), the discretization of the first equation of (\ref{AP3.6}) is a natural
choice for the first equation in the equilibrium diffusive radiative hydrodynamics (\ref{Ap2.9}).
This shows that the coupled GKS and UGKS method for the system (\ref{Ap2.1}) does have the asymptotic preserving property.
By virtue of (\ref{p2.3}), Eq.(\ref{p2.3.1}) becomes a standard five points scheme for the third (diffusion) equation of (\ref{Ap2.9}).
Therefore, the current scheme can capture the exact diffusion solution without the constraint on the cell size
being smaller than the photon's mean free path.

\hfill $\square$

\section{Numerical Results}\label{sec:results}

The coupled GKS and UGKS method will be tested in two radiative shock cases, which are presented in \cite{BHEM,Lowrie}.
For both shocks, the parameters are the monatomic gas $\gamma=5/3$, the specific heat $c_v=0.14472799784454$ $JK keV^{-1} g^{-1}(1 JK = 10^9 J)$,
the total absorption coefficient $\sigma_t=577.35 cm^{-1}$, and the scattering coefficient $\sigma_s=0$.
The specifications of the conditions in the far-stream pre and post-shock regions are provided in Table 1 and Table 2 for Mach $1.2$ and Mach $3$ shocks, respectively.
We initialize each radiative shock calculation by setting the states in the left half of the spatial domain
to be the far-stream pre-shock state and the states in the right half of the domain to be the post-shock state.
The CFL number is taken to be $0.6$ and the spatial cells are coarse mesh with $500$ points and fine mesh with $1000$ points.
The steady state solutions for both cases are obtained.

{\bf Example 1. (Mach $1.2$ shock)} First, for the weak  radiative shock at Mach $1.2$, the numerical results are shown in Fig. \ref{fig5.1}.
 There is a hydrodynamic shock, but no visible Zel'dovich spike \cite{Lowrie}.
 In the numerical solution, we observe a discontinuity in the fluid temperature due to the hydrodynamic shock,
 and the maximum temperature is bounded by the far-downstream temperature. This matches with the results in \cite{Lowrie, BHEM}.

{\bf Example 2. (Mach $3$ shock)} For the strong radiative shock at Mach $3$, the numerical results are shown Fig. \ref{fig5.2}.
 There have both a hydrodynamic shock and a Zel'dovich spike.
Discontinuities in both fluid density and temperature are observed in the hydrodynamic shock.
The Zel'dovich spike appears at the shock front with enhanced fluid temperature, and leads to a relaxation region downstream
where the fluid temperature and radiation temperature get to equilibrate. There is in good agreement with the results in \cite{Lowrie, BHEM}.


{\bf Example 3. (Interaction between a shock and a bubble)} This is about a Mach 3 shock, the same as in Example 2, interacting with a denser bubble.
Initially, there is a circular bubble of radius $R=0.2$ with its center located at (-0.008,0.01) in
the computation domain $[-0.02.0.4]2\times [0, 0.02]$. The bubble is $25$ times denser
than the surrounding gas, and the opacity parameter in the bubble is $100$ times of that in the ambient gas.
The shock is introduced at $x=-0.012$ with the same initial pre- and post-shock states for $(\rho, T, T_r, u)$
as given in Table 2, and the initial value for $v$ is zero. The upper and lower boundary conditions are zero gradients for
the flow variables and reflective for the radiation intensity. The radiation constant $a_R$ and the light speed are the same as given
in the first two examples. The computation is performed with $150\times 50$ cells. The final simulation time is taken to be $t=0.14ns$.
As a comparison, we also give the numerical solution of the Euler equations at the same output time.
As shown in the Fig. \ref{fig5.4}, the phenomena of Zel'dovich spike appears at the shock front by comparing
the material temperature with the radiation temperature at line $y=0.01$ in Fig. \ref{fig5.5}.
\begin{table}
\begin{center}
\begin{tabular}{llll}
\multicolumn{4}{c}{{{\small \bfseries Table 1 : Initial condition for the Mach 1.2 radiative shock problem }}
}\\
\hline
\itshape Parameter &\itshape Pre-shock Value &\itshape Post-shock Value &\itshape Units \\
\hline
$\rho$ &1.00000000e+00 &1.29731782e+00 & g cm$^{-3}$ \\
$u$    &1.52172533e-01 &1.17297805e-01 & cm $sh^{-1}$ \\
$T$    &1.00000000e-01 &1.19475741e-01 & keV \\
$E_r$  &1.37201720e-06 &2.79562228e-06 & Jk cm$^{-3}$ \\
\hline
\end{tabular}
\end{center}
\end{table}

\vspace{2mm}

\begin{table}
\begin{center}
\begin{tabular}{llll}
\multicolumn{4}{c}{{\small \bfseries Table 2 : Initial condition for the Mach 3 radiative shock problem}}\\
\hline
\itshape Parameter &\itshape Pre-shock Value &\itshape Post-shock Value &\itshape Units \\
\hline
$\rho$ &1.00000000e+00 &3.00185103e+00 & g cm$^{-3}$ \\
$u$    &3.80431331e-01 &1.26732249e-01 & cm $sh^{-1}$ \\
$T$    &1.00000000e-01 &3.66260705e-01 & keV \\
$E_r$  &1.37201720e-06 &2.46899872e-04 & Jk cm$^{-3}$ \\
\hline
\end{tabular}
\end{center}
\end{table}

\vspace{1mm}

\begin{figure}
\centering
%
\rotatebox{0}{\includegraphics[width=7cm]{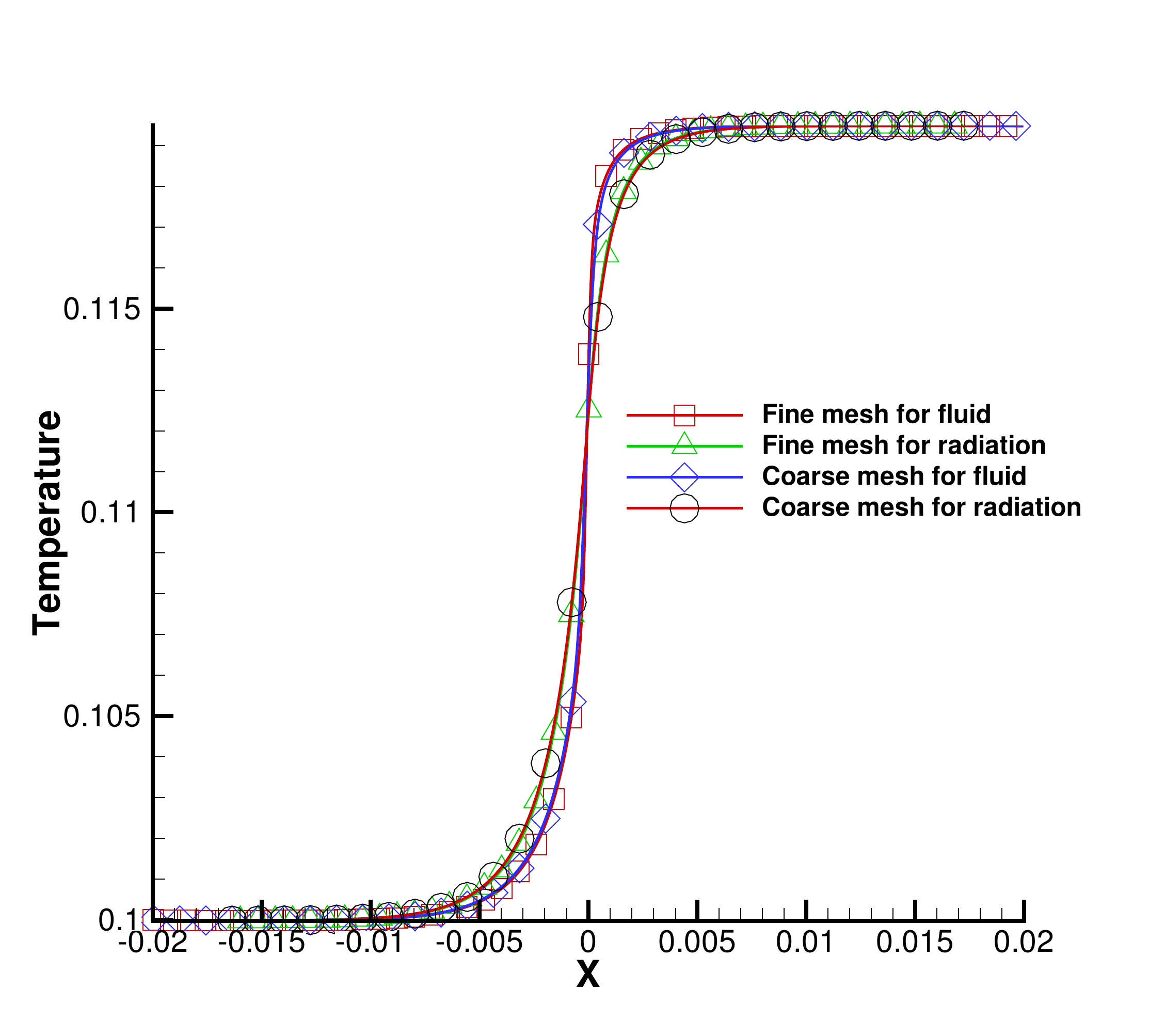}} \\
\rotatebox{0}{\includegraphics[width=7cm]{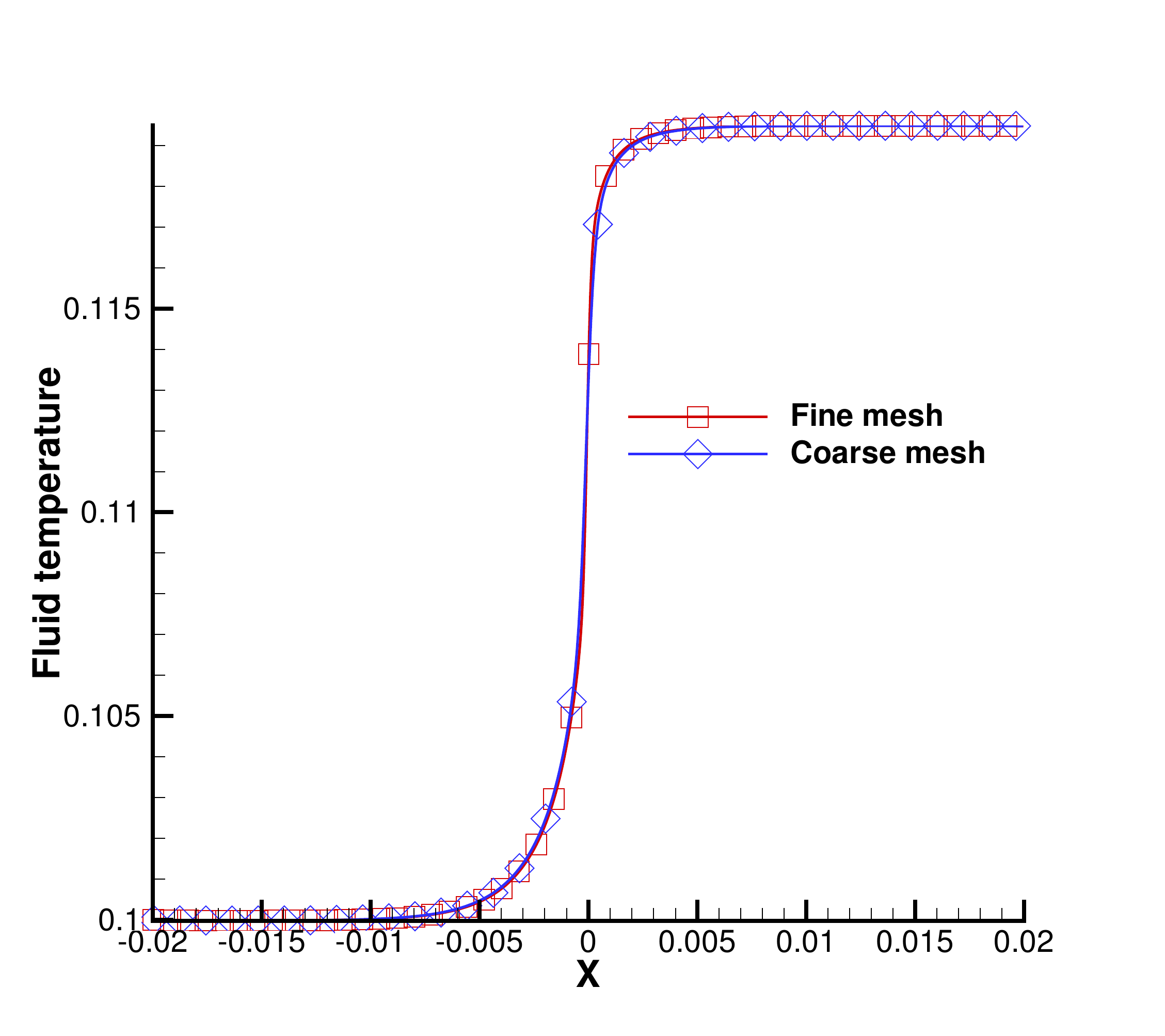}} \ \
\rotatebox{0}{\includegraphics[width=7cm]{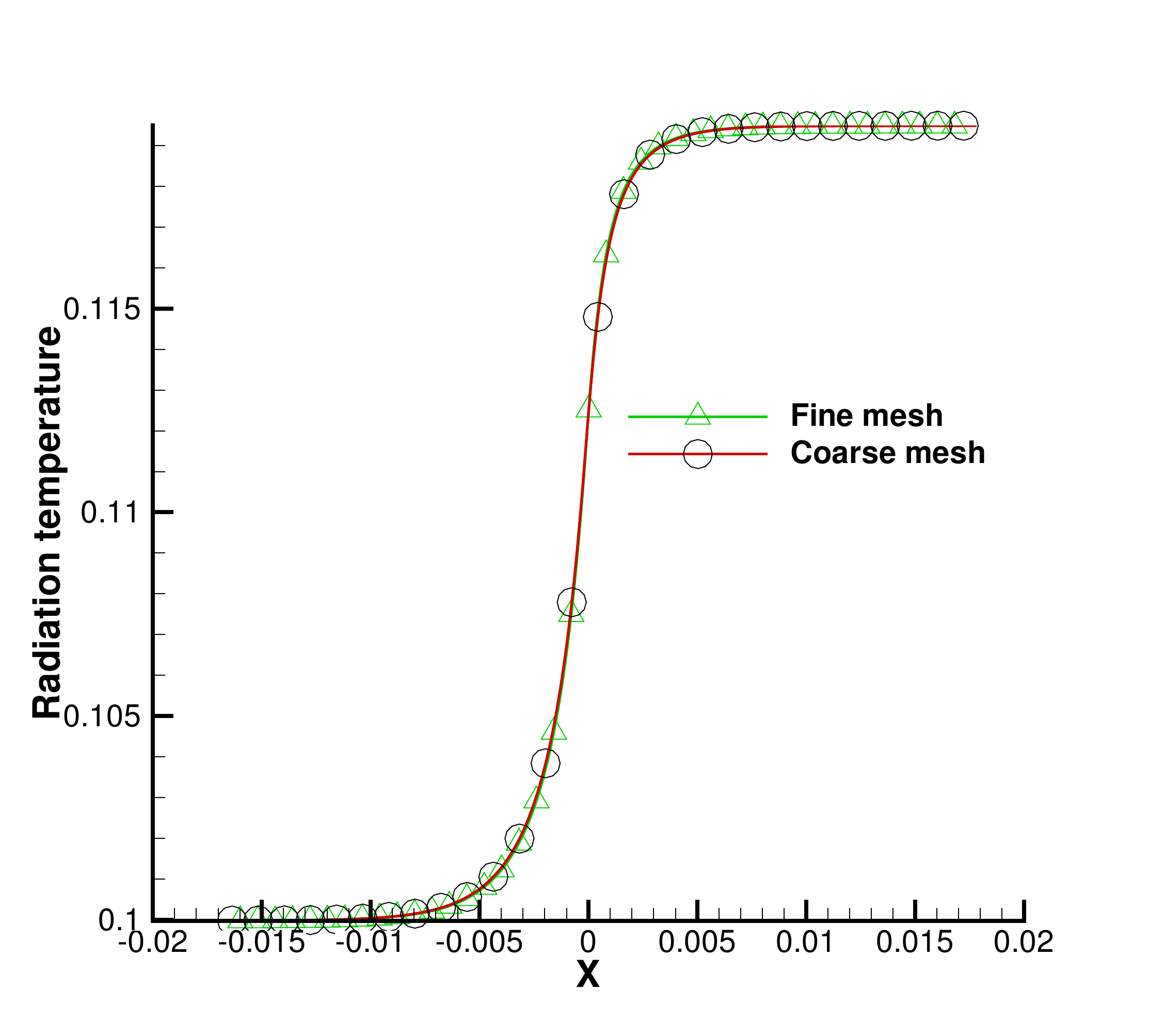}} \\
\rotatebox{0}{\includegraphics[width=7cm]{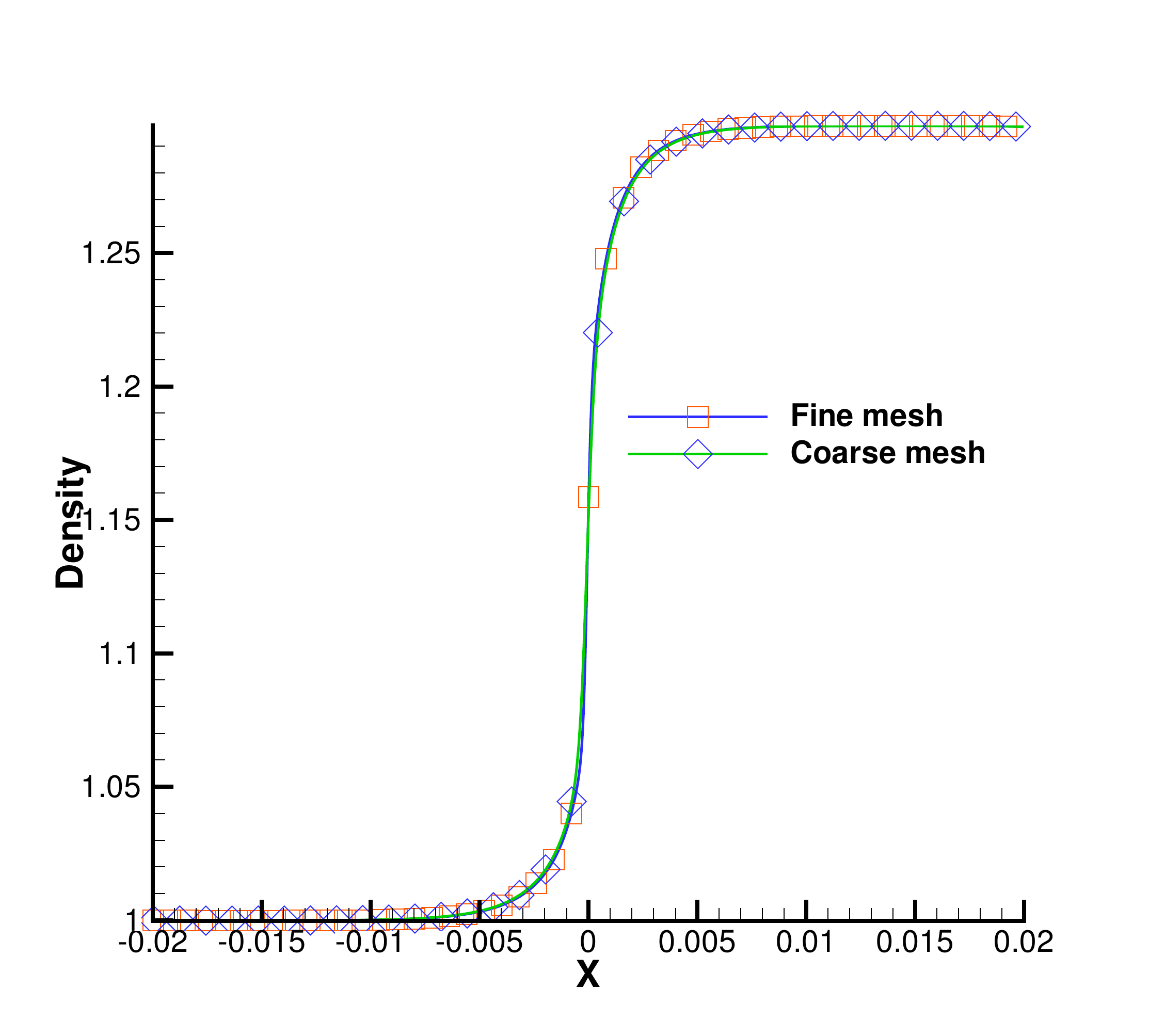}} \ \
\rotatebox{0}{\includegraphics[width=7cm]{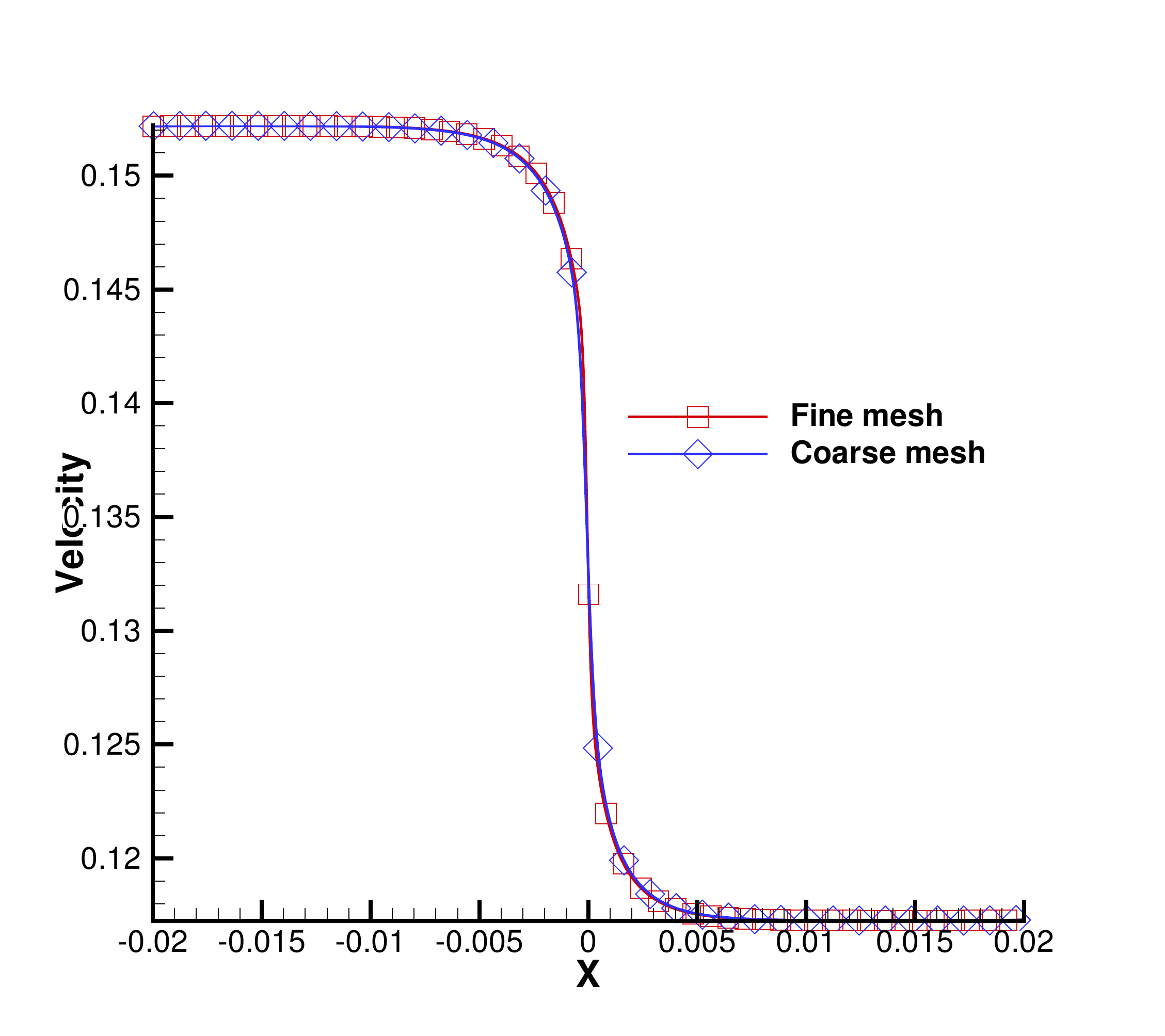}}

\caption{\label{fig5.1}{{\small Numerical results for Mach 1.2 radiative shock.} } }
\end{figure}

\vspace{1mm}

\begin{figure}
\centering
%
\rotatebox{0}{\includegraphics[width=7cm]{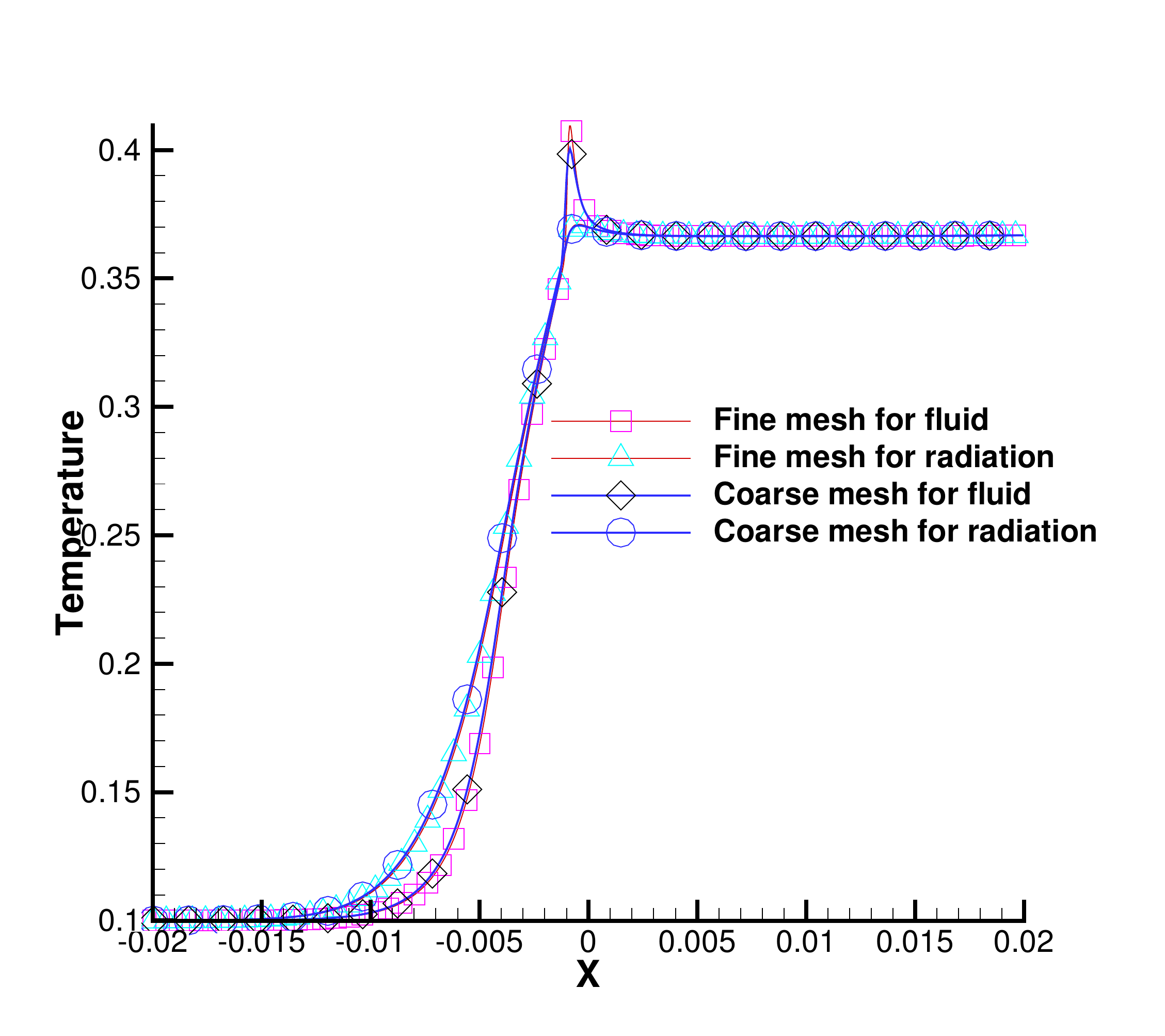}} \\
\rotatebox{0}{\includegraphics[width=7cm]{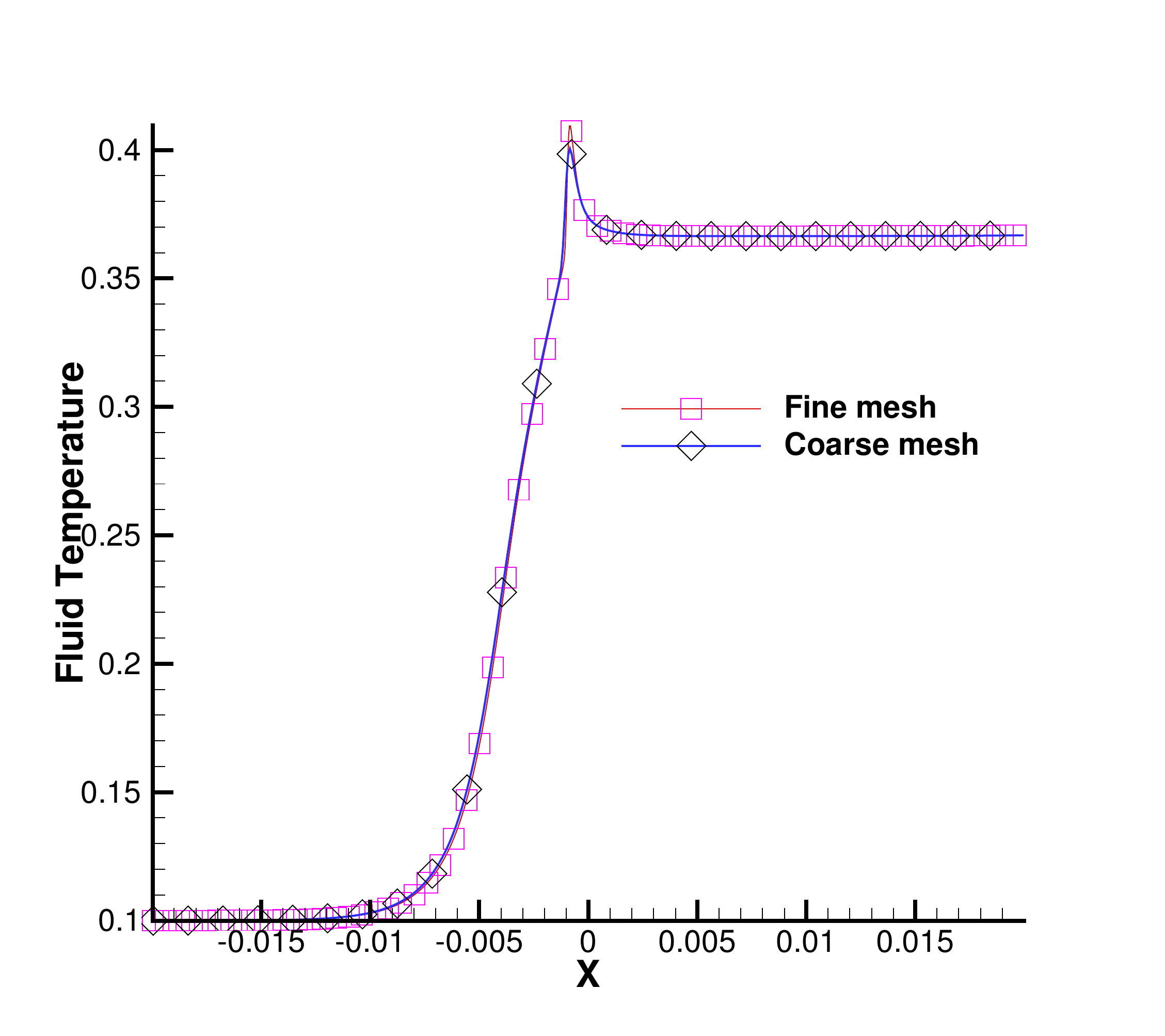}} \ \
\rotatebox{0}{\includegraphics[width=7cm]{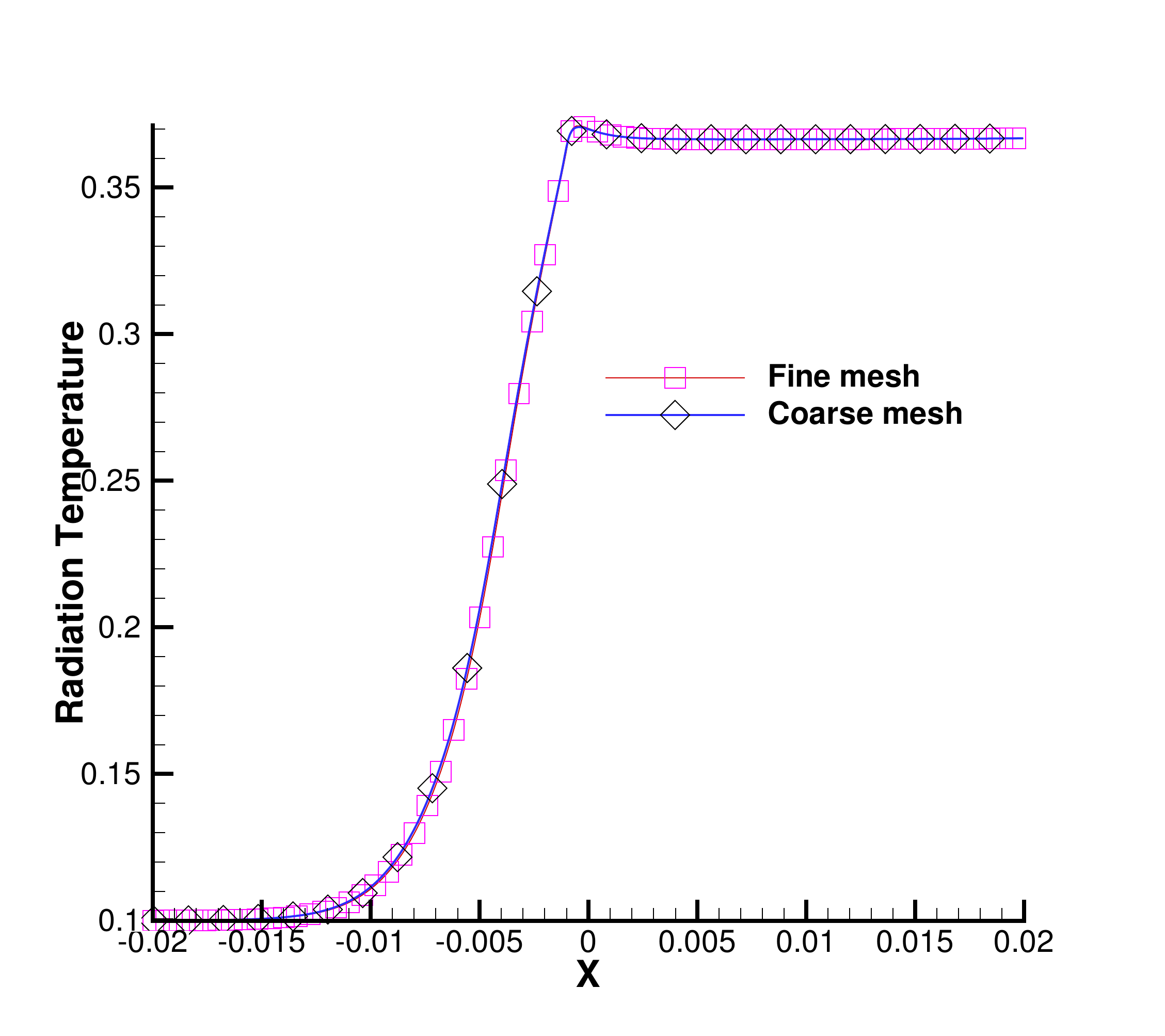}} \\
\rotatebox{0}{\includegraphics[width=7cm]{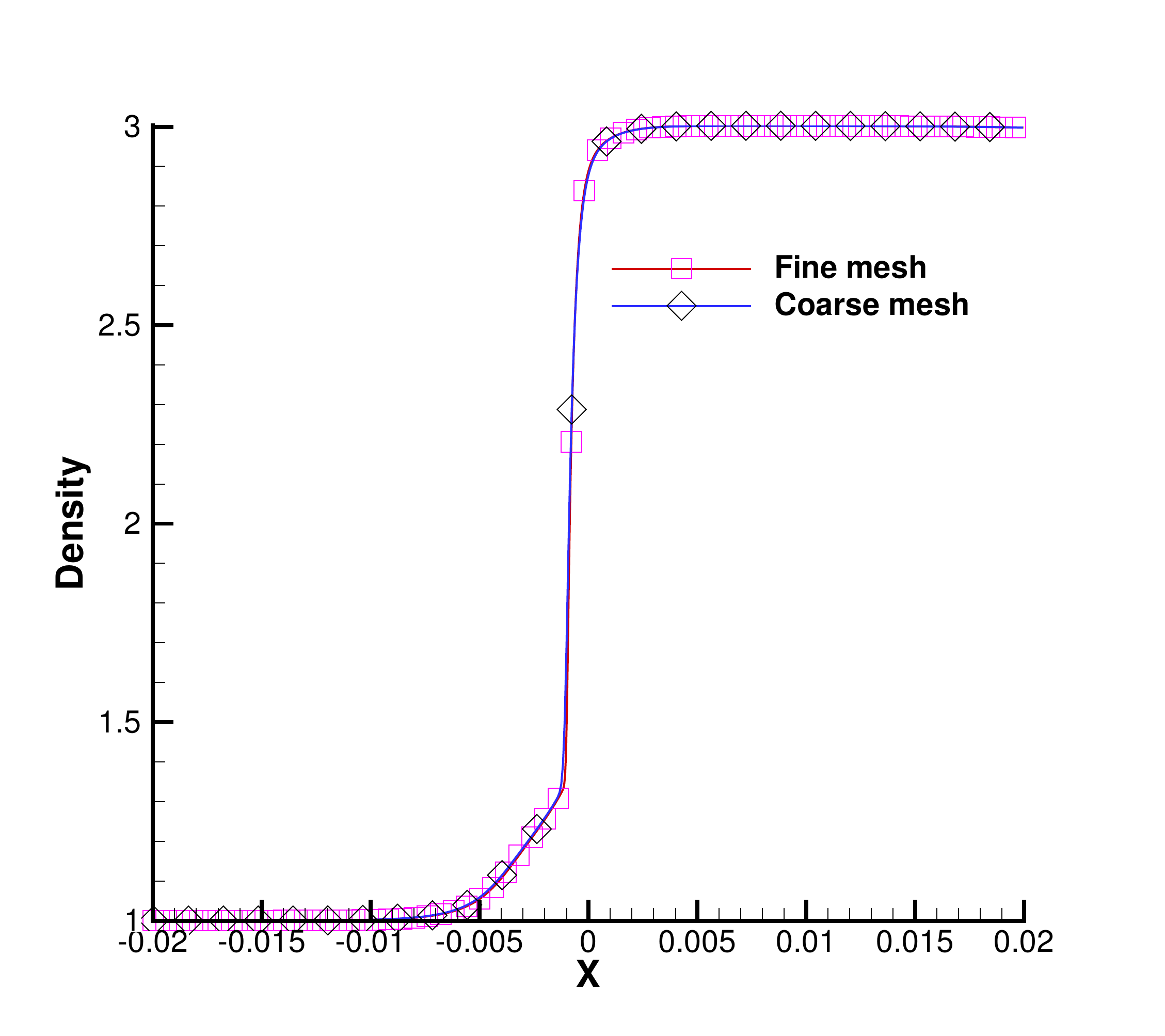}} \ \
\rotatebox{0}{\includegraphics[width=7cm]{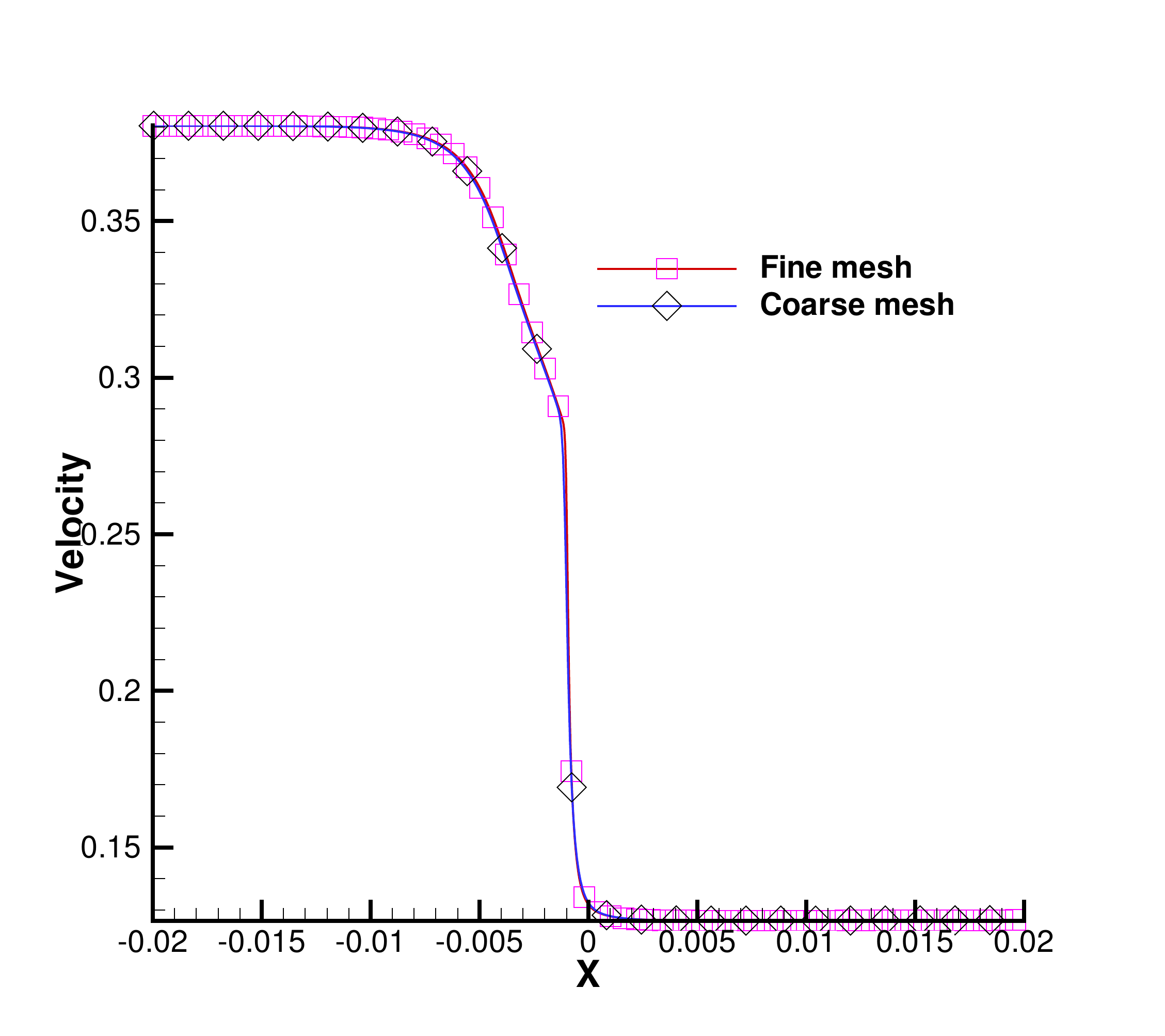}}  \\

\caption{\label{fig5.2}{{\small Numerical results for Mach 3 radiative shock.} } }
\end{figure}

\vspace{1mm}

%


\begin{figure}
\centering
%
\rotatebox{0}{\includegraphics[width=6cm]{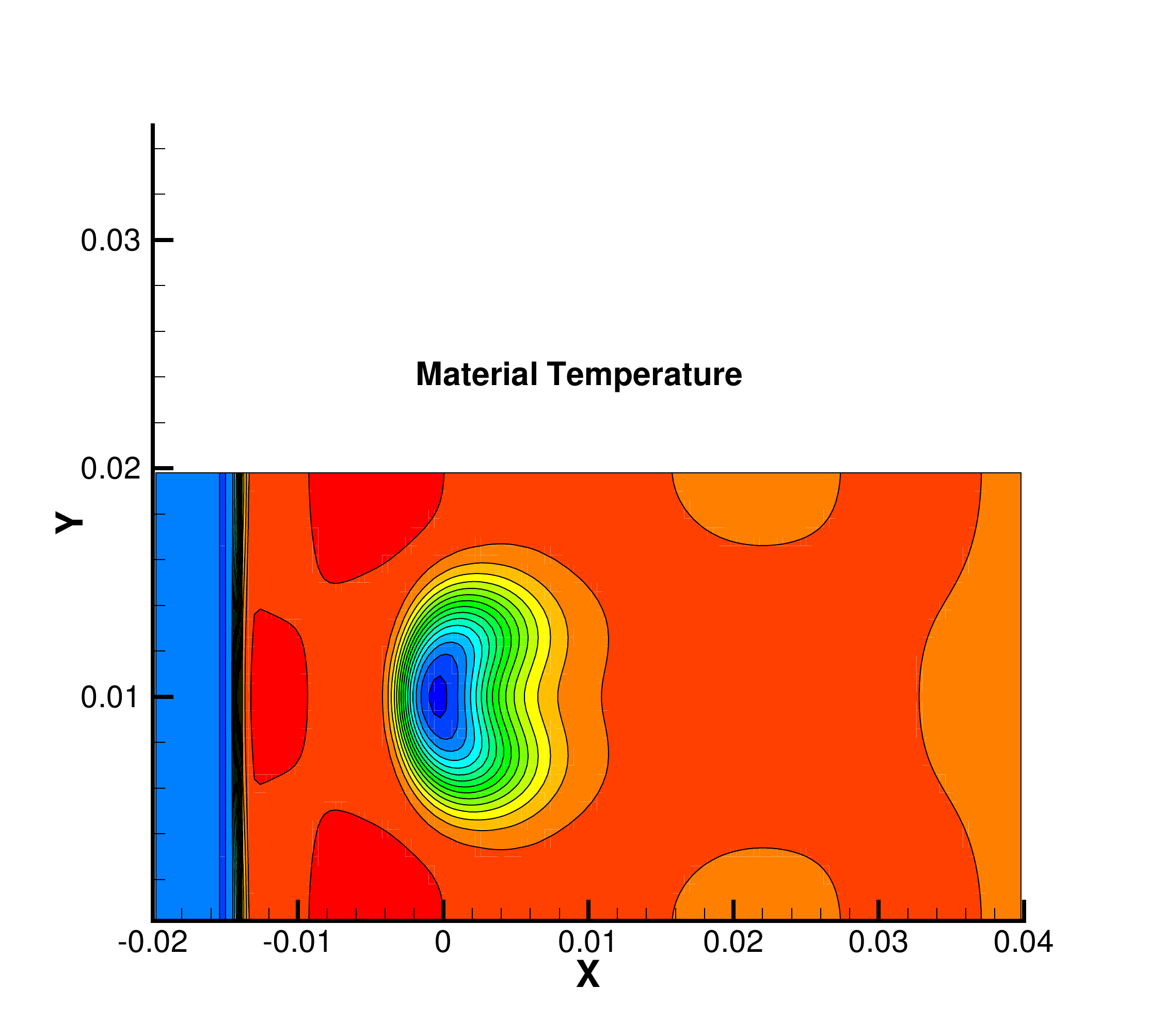}} \ \
\rotatebox{0}{\includegraphics[width=6cm]{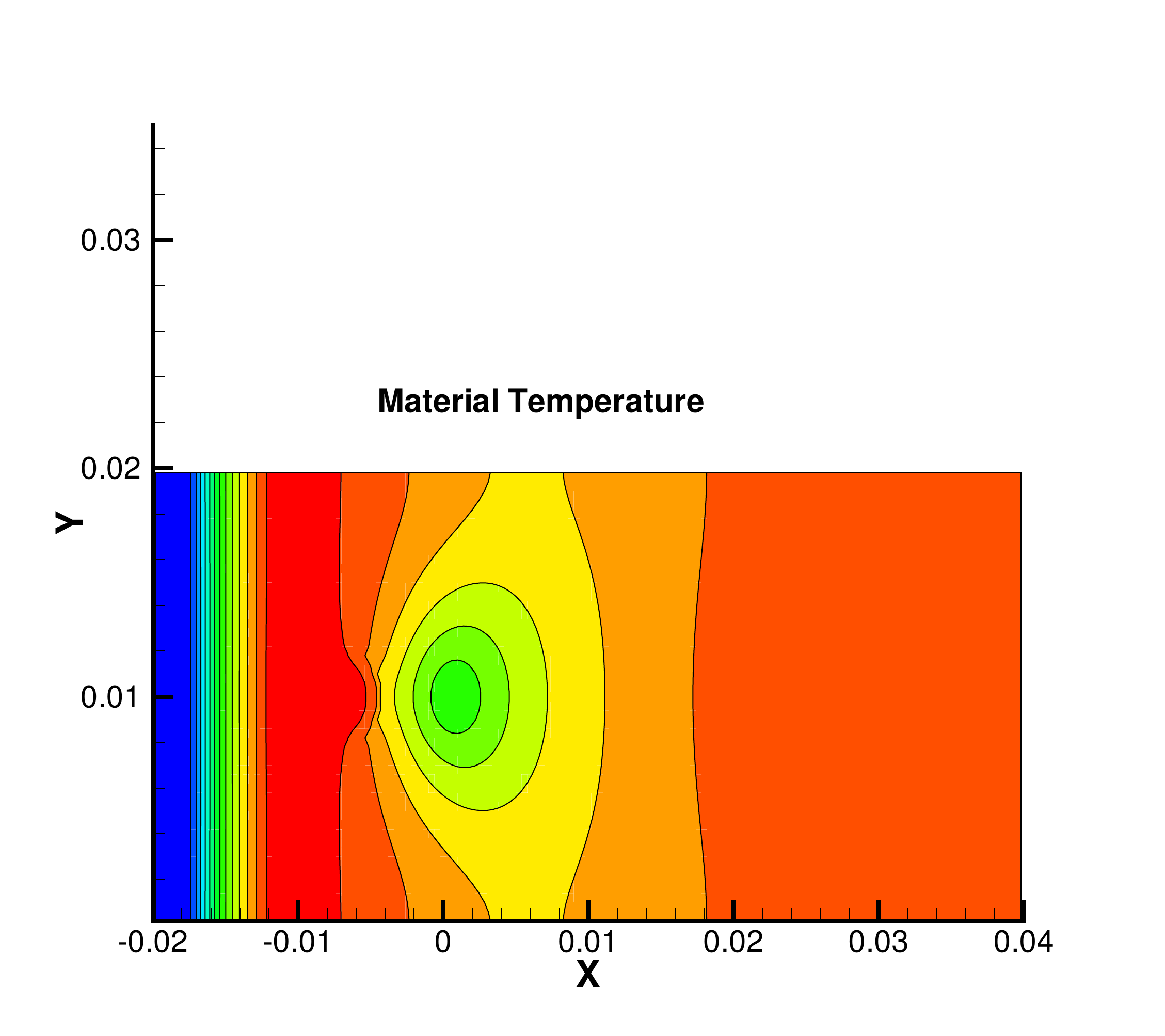}} \\
\rotatebox{0}{\includegraphics[width=6cm]{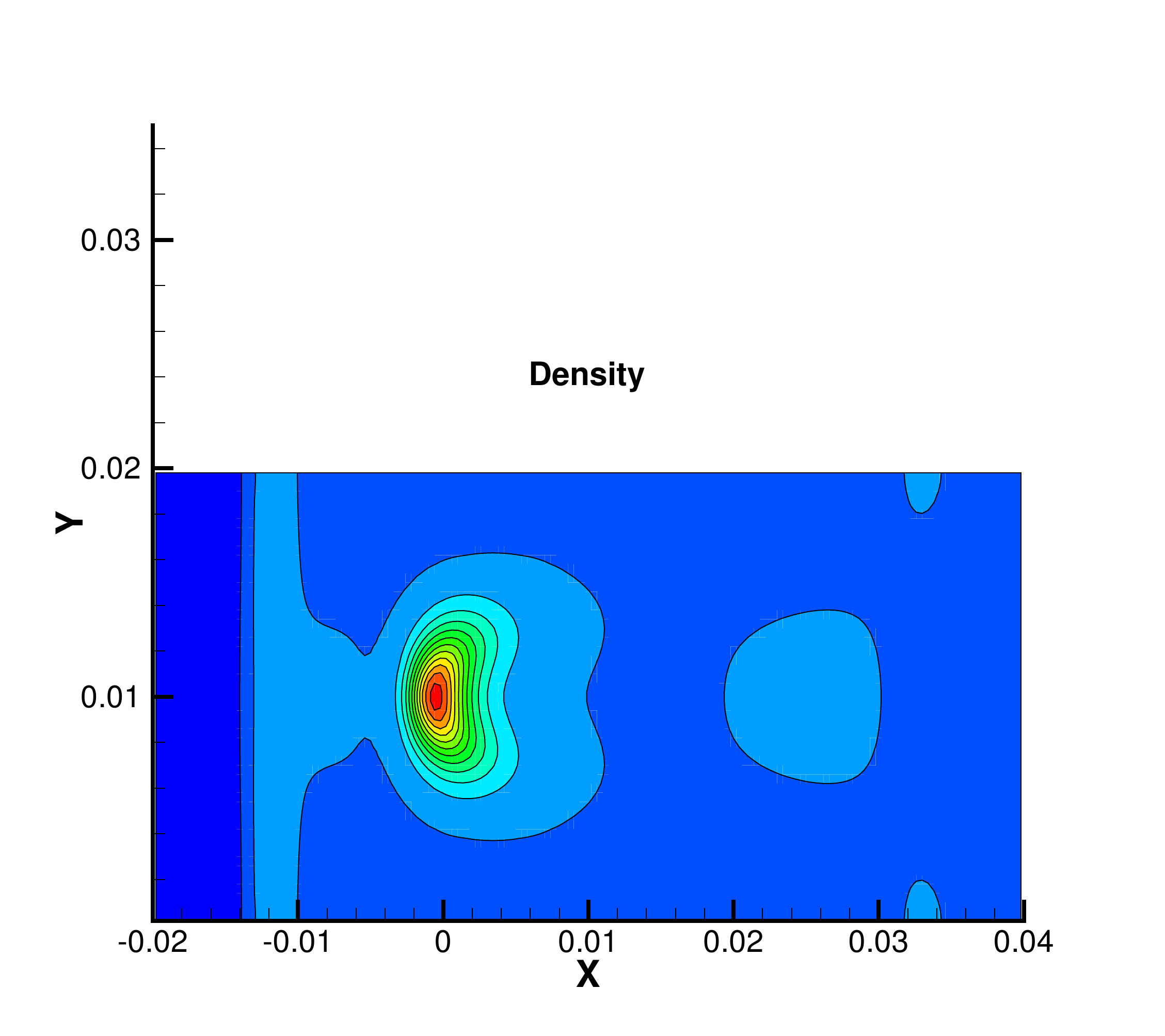}} \ \
\rotatebox{0}{\includegraphics[width=6cm]{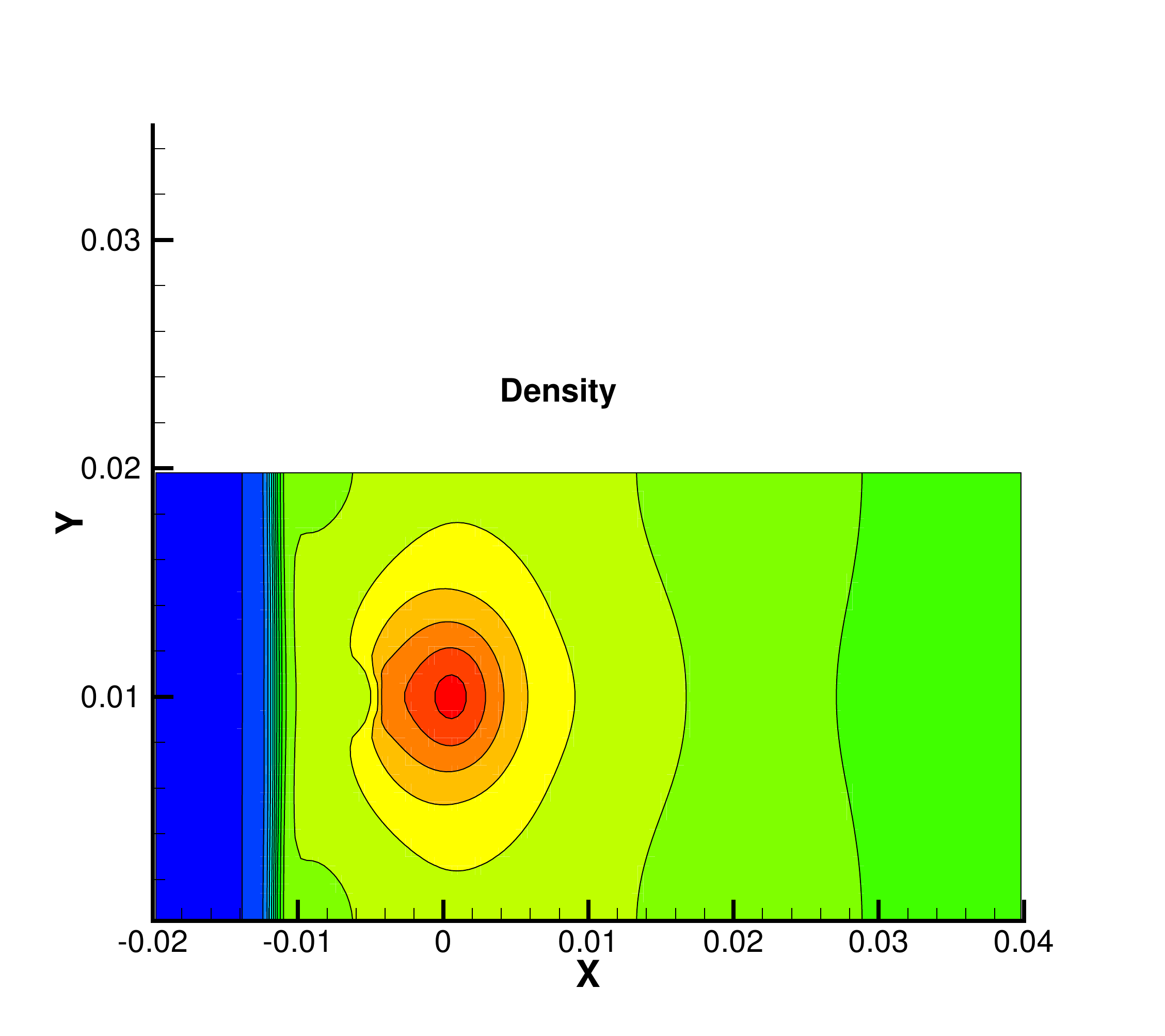}} \\
\rotatebox{0}{\includegraphics[width=6cm]{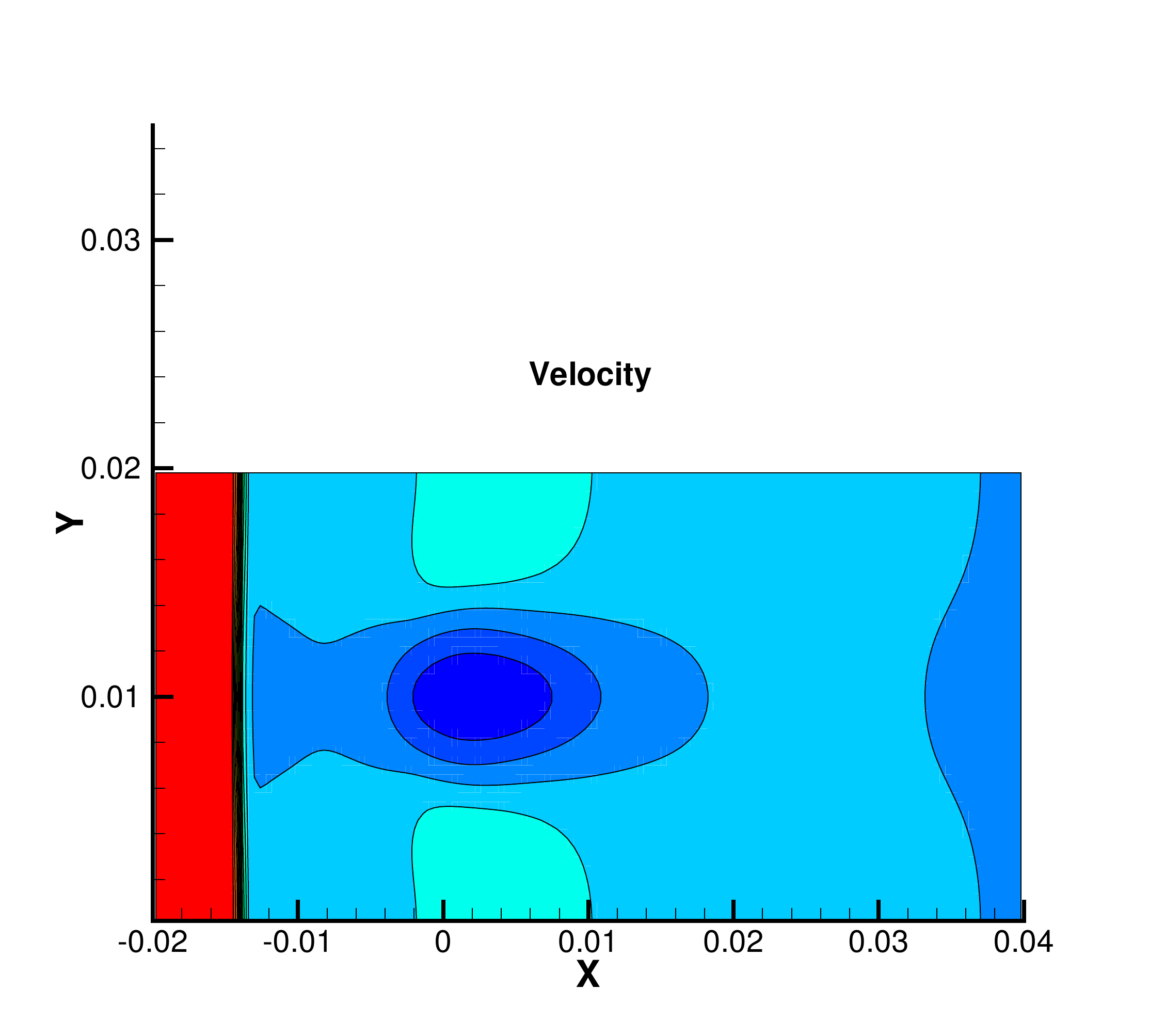}}  \ \
\rotatebox{0}{\includegraphics[width=6cm]{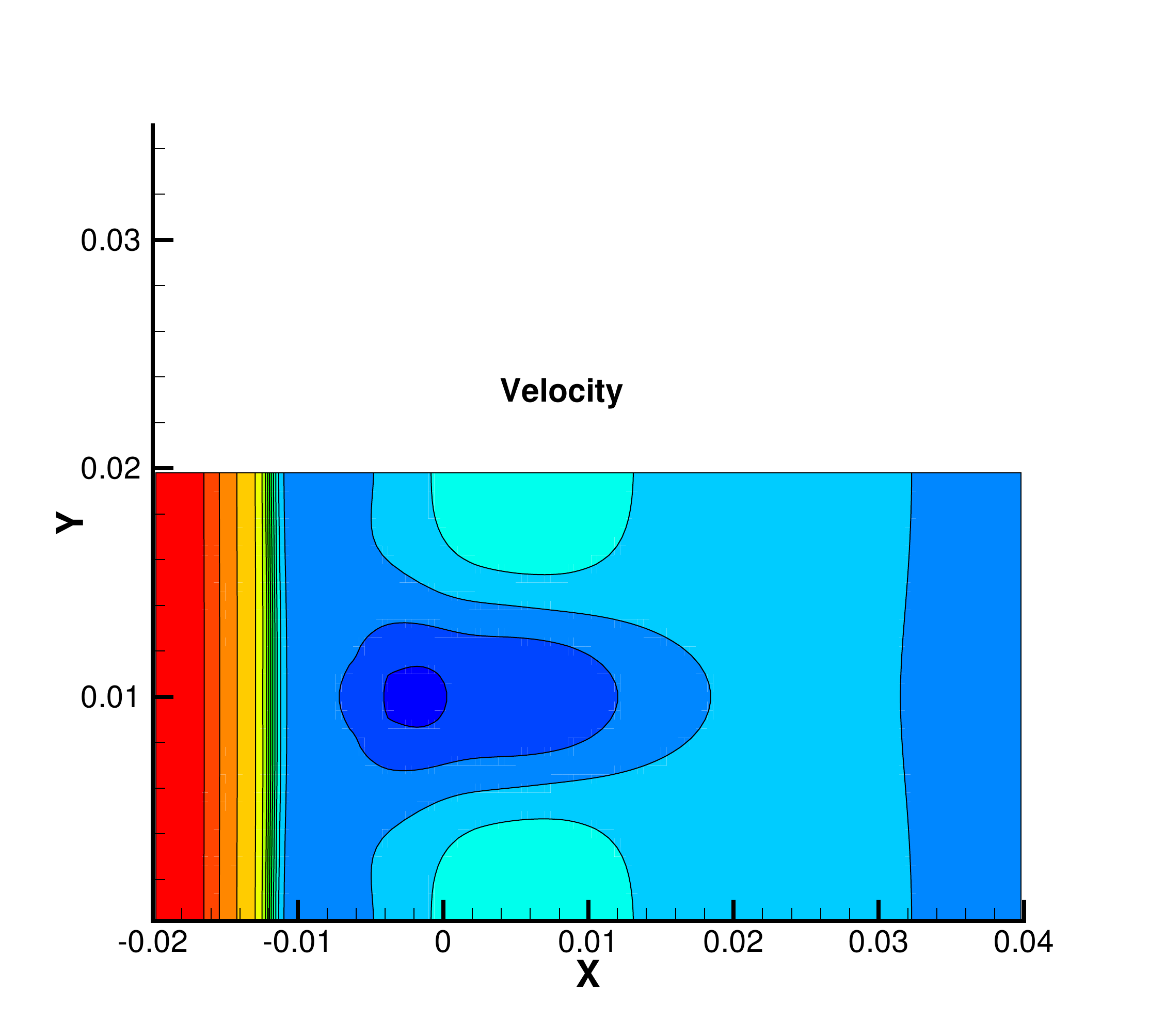}} \\
\rotatebox{0}{\includegraphics[width=6cm]{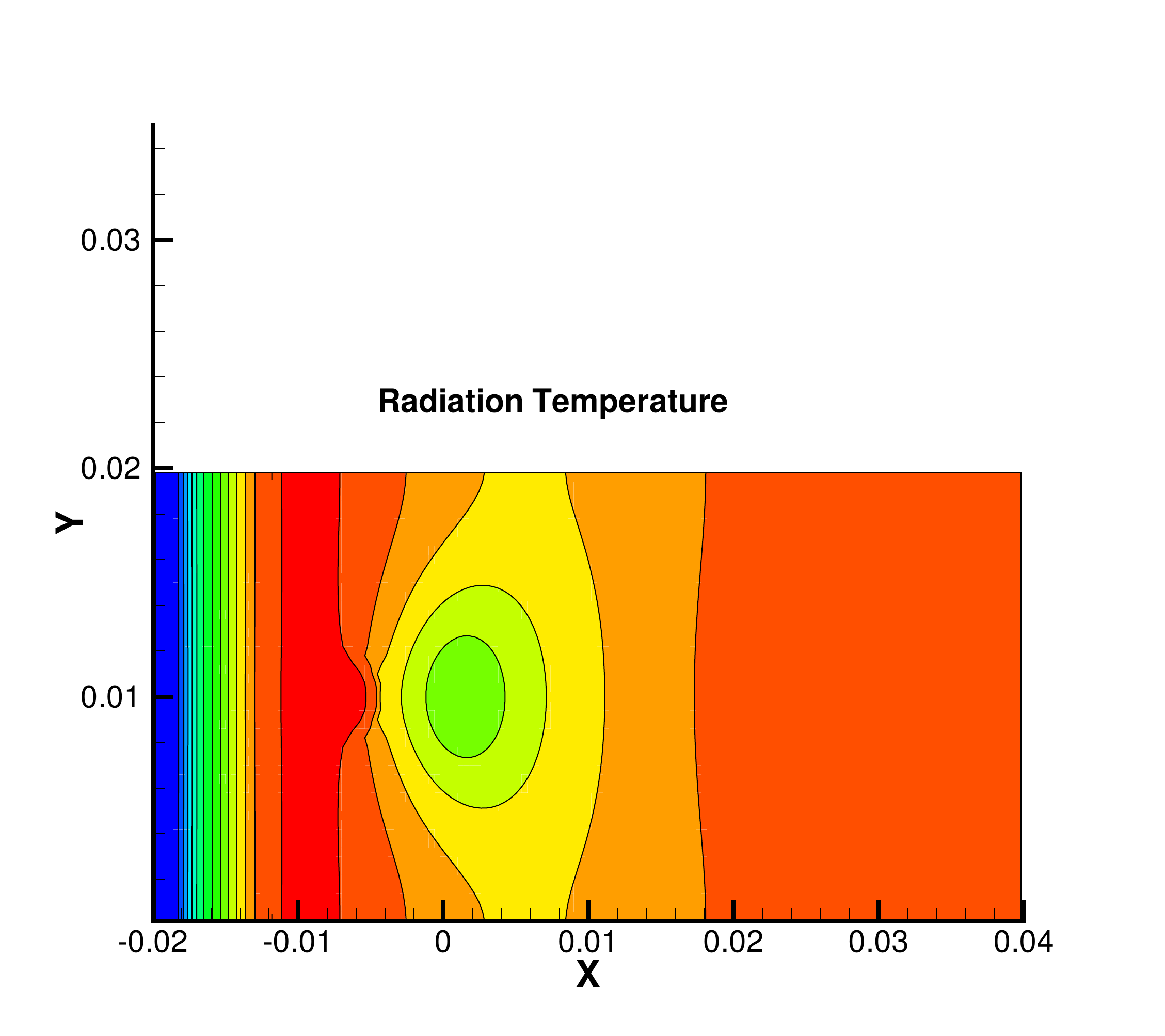}}
\caption{\label{fig5.4}{{\small Computed results at time $t=0.14ns$. The left figures are the temperature, density, and velocity of the purely
Euler gas dynamic solution, while the right figures are the numerical results for radiation hydrodynamics.
The lower one is the computed radiation temperature.} } }
\end{figure}

\vspace{1mm}

\begin{figure}
\centering
%
\rotatebox{0}{\includegraphics[width=8cm]{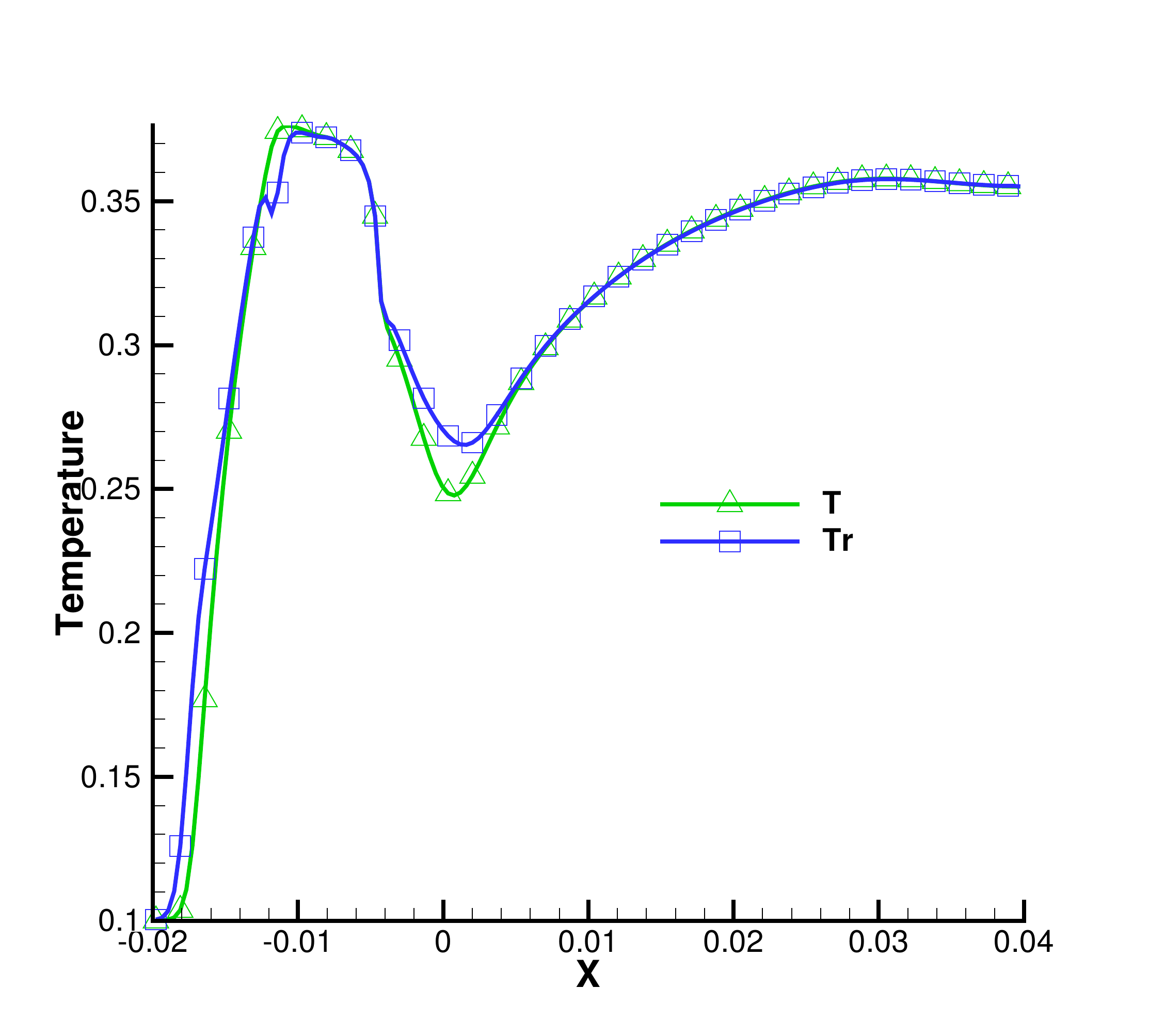}}
\caption{\label{fig5.5}{{\small Computed material and radiation temperatures at line $y=0.01$ and time $t=0.14ns$ in Example 3 } } }
\end{figure}

\section{Conclusion}\label{sec:con}

In this paper, we have presented a coupled GKS and UGKS for the numerical solution of the equations of radiation hydrodynamics.
The model equations consist of the fluid dynamic equations and a gray radiative transfer equation, with the momentum and
energy exchange between them. For the hydrodynamic part, the GKS is used to solve the compressible flow equations,
while for the radiative transfer part, the multiscale UGKS is adopted. Since both GKS and UGKS are finite volume methods,
all unknowns are defined inside each cell and consistent discretizations for the hydrodynamics and radiative transfer can be constructed.
Due to the multiscale nature of UGKS, the final scheme has the asymptotic preserving property for the whole system.
The coupled scheme can recover the equilibrium diffusion limit of radiation hydrodynamic equations in the optically thick region, and
has no requirement of cell size being less than photon's mean free path.
The standard radiation shock wave problems and the interaction between the shock and dense bubble have been tested
to validate the proposed scheme.

\section*{Acknowledgement} 
The current research  is supported by NSFC (Grant Nos. 11671048, 91630310) for Sun;
by NSFC (Grant Nos. 11631008, GZ1465, 11571046) for Jiang; and by Hong Kong research grant council (16206617)
and  NSFC (Grant Nos. 11772281,91852114) for Xu.



\clearpage
%


\end{document}